\documentclass[aps,prx,reprint,groupedaddress,showpacs]{revtex4-1}

\usepackage{graphicx} 
\usepackage{dcolumn}
\usepackage{amsfonts}
\usepackage{amssymb}
\usepackage{amsmath}
\usepackage{bm}
\usepackage{epstopdf}
\newcommand{\be}{\begin{equation}}
\newcommand{\ee}{\end{equation}}
\newcommand{\bea}{\begin{eqnarray}}
\newcommand{\eea}{\end{eqnarray}}

\renewcommand{\vec}[1]{{\mathbf #1}}

\begin{document} 

\title{Imaging topologically protected transport with quantum degenerate gases}

\author{Brian Dellabetta$^{1,2}$}
\author{Taylor L. Hughes$^3$}
\author{Matthew J. Gilbert$^{1,2}$}
\author{Benjamin L. Lev$^4$}
\affiliation{$^1$Department of Electrical and Computer Engineering, University of Illinois, Urbana, IL 61801}
\affiliation{$^2$Micro and Nanotechnology Laboratory, University of Illinois, Urbana, IL 61801}
\affiliation{$^3$Department of Physics, University of Illinois, Urbana, IL 61801}
\affiliation{$^4$Departments of Applied Physics and Physics and Ginzton Laboratory, Stanford University, Stanford, CA 94305}

\begin{abstract}
Ultracold and quantum degenerate gases held near conductive surfaces can serve as sensitive, high resolution, and wide-area probes of electronic current flow.  Previous work has imaged transport around grain boundaries in a gold wire by using ultracold and Bose-Einstein condensed atoms held microns from the surface with an atom chip trap.  We show that atom chip microscopy may be  applied to useful purpose in the context of materials exhibiting topologically protected surface transport.  Current flow through lithographically tailored surface defects in topological insulators (TI)---both idealized and with the band-structure and conductivity typical of Bi$_{2}$Se$_{3}$---is numerically calculated.  We propose that imaging current flow patterns enables the differentiation of an ideal TI from one with a finite bulk--to--surface conductivity ratio, and specifically, that the determination of this ratio may be possible by imaging transport around trenches etched into the TI's surface. 
\end{abstract}
\date{\today}
\maketitle

\section{Introduction}

The topological insulator is a unique state of topologically non-trivial quantum matter that has sparked a tremendous amount of interest in the condensed matter community~\cite{Hasan10,Qi11}. Interest is not only focused on the novel manner in which this matter organizes, which is distinct from the standard Landau symmetry-breaking paradigm, but also on the potential use of topological insulators in spintronic devices and in topologically protected quantum information processing~\cite{Fu:2007,Moore:2007,Qi:2008,Schnyder:2008,Flensberg11}.  The former application may arise from the strong spin-momentum locking of electron transport in the surface state.  In heterostructures formed with, i.e., superconductors or topological superconductors, novel electronic excitations, particularly Majorana fermions, may arise whose manipulation (braiding) is thought to provide a means with which to engineer topologically protected quantum computation~\cite{Hasan10,Qi11}. 

Unfortunately, all known topological insulator materials suffer from large bulk conduction:  they are not truly insulating due to chemical imperfections~\cite{Qi11} and so are not readily amenable to traditional transport measurements.  For example, in the ternary chalcogenide Bi$_{2}$Te$_{2}$Se, the surface--to--bulk conductivity ratio is $\sim$6\%~\cite{Ando10}, though the measurements on the exact ratio  differ~\cite{Ong11}.  This makes surface transport properties very difficult to measure, let alone manipulate~\cite{Fisher10,JarilloHerrero11}.  A  challenge lies in acquiring the ability to distinguish, via transport, the interesting surface state dynamics from the less interesting dynamics in the bulk.   Only then will the promise of novel electronic devices, exotic quantum phenomena, and quantum information processors be realized with topological insulators.

While the existence of the topologically protected surface state has been unambiguously detected in angle-resolved photoemission spectroscopy (ARPES)~\cite{Chen:2009,Hsieh:2009} and scanning tunneling microcopy (STM)~\cite{Roushan:2009,Alpichshev:2011,Yazdani11} experiments, none of these techniques directly probe surface transport, which the aforementioned applications rely upon for functionality.  But more fundamentally, there is significant disagreement~\cite{Analytis12,Butch10} about the nature of the surface state itself due to contradictory measurements from the disparate techniques of APRES, STM, quantum oscillations, and Hall conductance measurements~\cite{Ando09,Fisher10,Ando10,Ong10QO,Ong10,Butch10}.   Band structure may bend at surfaces, inducing a crossing of the Fermi energy only at the surface, and surface probes, such as ARPES and STM, may then give a skewed picture of the material as a whole~\cite{Analytis10,Butch10}.  Conduction band states in the doped bulk form a parallel conducting path that cannot be effectively removed by electrostatic gating~\cite{Steinberg:2010} and traditional transport and Hall measurements on samples of varying geometry require a number of assumptions to analyze transport data, even on nanosamples~\cite{JarilloHerrero11,Analytis12}.    Time-resolved fundamental and second harmonic optical pump-probe spectroscopy can reveal differences in transient responses in the surface versus bulk states~\cite{Gedik11}, but this detection method does not isolate transport properties in the surface from the bulk.   

Extant methods are not wholly satisfying from the standpoint of robustly detecting topologically protected surface states in presumptive topological insulators in a relatively model-independent fashion.  By contrast, this proposal  presents an independent technique that enables the direct detection of the surface current in a manner that provides a relatively model-free measure of the surface conductivity versus the bulk conductivity.  This information may prove crucial in attempts to improve material growth techniques for obtaining more ideal topological insulators, such as those amenable to the spintronic and topologically protected quantum information processing applications mentioned above.

The atom chip microscope presented here may also advance topological insulator physics in other manners.  The surface state of existing topological insulators seems to be fragile in that over time and exposure, APRES and terahertz spectroscopy have shown modifications to the surface and bulk states~\cite{Cui11,Armitage12}.  Such aging effects will hamper device functionality unless fabrication techniques mitigate this effect.  A surface transport probe such as we propose should be a powerful tool to diagnose these aging effects under various preparation conditions.  Moreover, dynamically adding magnetic impurities to the surface breaks time-reversal symmetry in a way that should disrupt surface transport of the Dirac state~\cite{Hasan11,Analytis12}.  The cryogenic atom chip microcope would be well-poised to observe such dynamics.

 Taking a long-term perspective, the proposed microscope serves a dual purpose in that it may enable the coherent coupling of matter waves to Majorana fermions, for either imaging or for building topologically protected quantum hybrid circuits~\cite{Flensberg11}.  Indeed, the atom chip microscope may serve as an interesting probe of transport in topological superconducting systems of copper intercalated Bi$_2$Se$_3$~\cite{Cava10,Ando11}.

In this work, we examine the possibility of utilizing magnetic field signatures from electronic transport in TIs as a means of characterizing the topologically protected surface state.   Specifically, we propose  the use of atom chips---substrates supporting micron-sized current-carrying wires that create magnetic microtraps near surfaces for thermal gases or Bose-Einstein condensates (BECs)---to enable single-shot and raster-scanned large-field-of-view detection of magnetic fields emanating from electronic transport in a TI.  A previous proposal noted the utility of atom chip microscopy in the context of imaging transport in two-dimensional electron gases and employing these electron gases for atom trapping~\cite{Fromhold11}. 

\begin{figure}[t]
\includegraphics[width=0.4\textwidth]{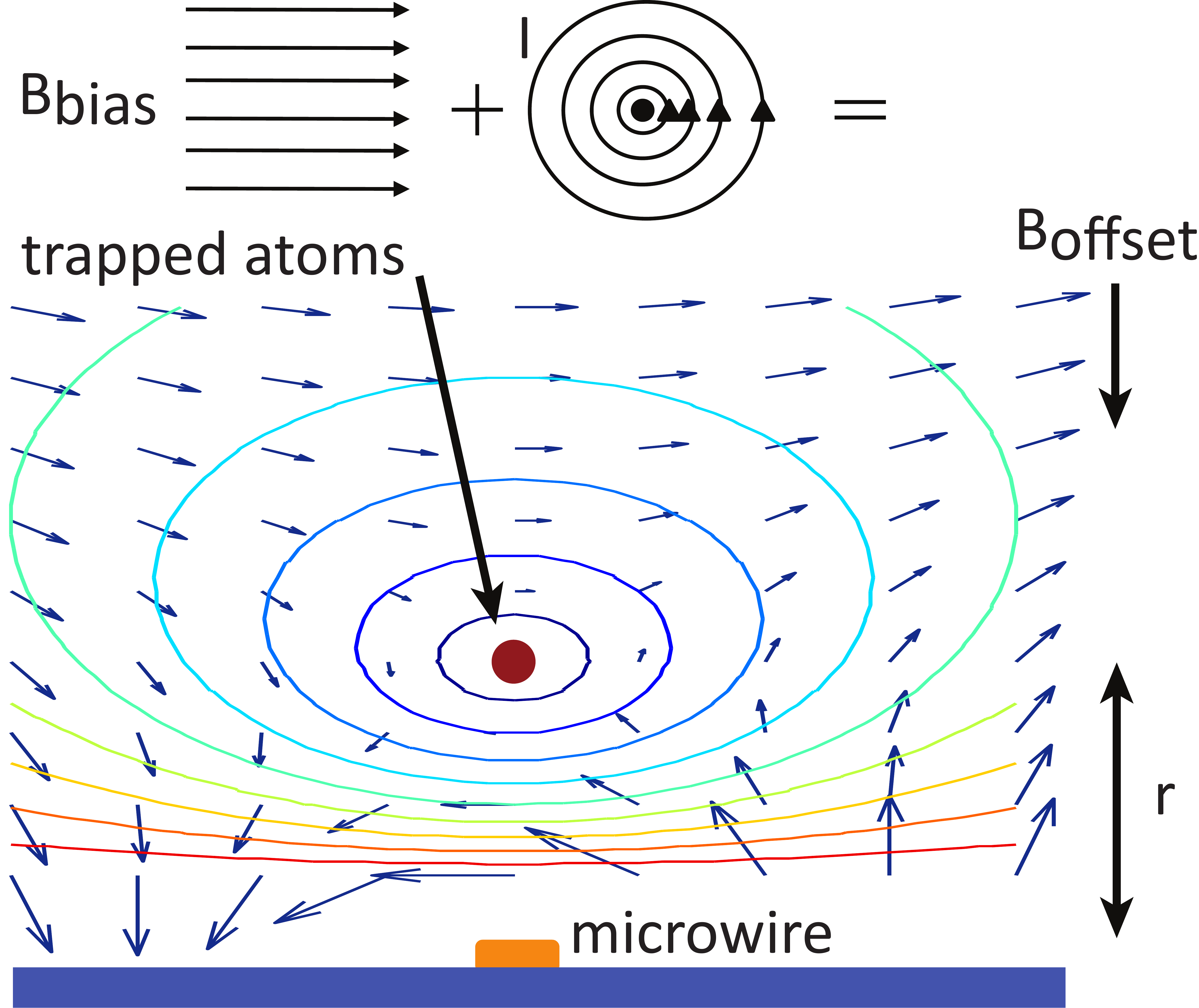}
\caption{(color online).    Atom chip trapping principle~\cite{Reichel01a,Schmiedmayer02,Zimmermann07}.  A cylindrically symmetric quadrupole magnetic field is created by superimposing a weak, homogeneous bias field $B_\text{Bias}$ with that from a wire with current $I$.  Weak-field seeking atoms---i.e., atoms in a Zeeman state whose energy increases for increasing magnetic field magnitude---are trapped in a small region around the zero of the magnetic quadrupole field.  If the microwire is attached to a surface via standard photolithography, then the trap may be brought to an arbitrarily close distance $r\propto I/B_\text{Bias}$ from the surface by adjusting the ratio of $I$ to $B_\text{Bias}$.  (Surface potentials limit $r>200$ nm.) Trapped atoms may be translated perpendicular to the axis of the wire by rotating with an offset field $B_\text{Offset}$ the angle that $B_\text{Bias}$ subtends with the substrate surface, thus enabling the precise positioning of atoms above adjacent materials.}
\label{fig:atomchip}
\end{figure}

\begin{figure*}[t]
\includegraphics[width=1\textwidth]{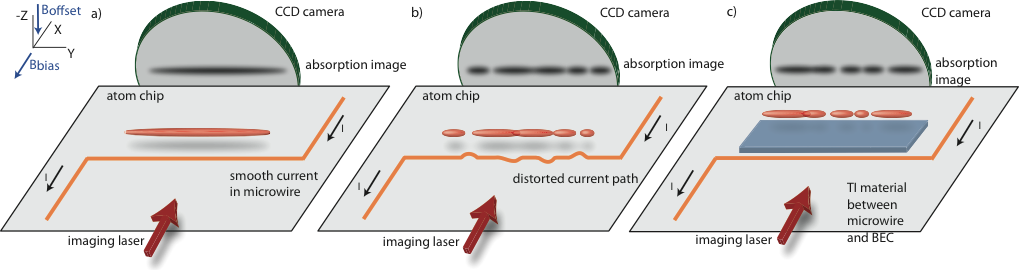}
\caption{(color online).  Atom chip microscopy principle~\cite{Schmiedmayer05_Nature,Schmiedmayer06_APL}.  (a)  A Bose-Einstein condensate (BEC, shown in red) is confined above the atom chip by the Z-trap formed by the fields from the Z-shaped microwire (orange) and $B_\text{Bias}$ produced by an external (not shown) Helmholtz coil pair whose axis is aligned with $\hat{x}$.  Adjusting current $I$ in the trapping wire and $B_\text{Bias}$ controls---with sub-micron precision---the position of the BEC above the surface of the substrate.  (b) Current running through the sample wire may not flow parallel to the wire's axis due to scattering centers (exaggerated in figure), which result in 1 ppm variations of the magnetic field above the sample~\cite{Aigner:2008}.  This inhomogeneous field deforms the trap, imprinting density modulations onto the otherwise smooth, cigar-shaped BEC cloud.  The absorption of a near-infrared laser casts a shadow onto a CCD camera, providing $\mu$m-scale resolution of density perturbations in the sub-micron wide, $\sim$1-mm-long cloud~\cite{Kruger07}.  The BEC can be recreated and repositioned every few seconds, thus in a few minutes providing a wide area map of the inhomogeneous current flow.  (c) To image transport in a TI, the atoms can be cantilevered over the sample---and held away from the Z-wire---by rotating $B_\text{Bias}$ in the $xz$-plane using $B_\text{Offset}$.  The TI sample (shown in blue) may be mounted on the cryogenically cooled atom chip.}
\label{fig:ACM}
\end{figure*}

Figures~\ref{fig:atomchip} and~\ref{fig:ACM} depict the principles of atom chip trapping and the atom chip microscope, respectively. (The figure captions contain the descriptions of the geometry and measurement scheme.)  Cryogenic atom chip microscopy introduces very important features to the toolbox of high-resolution, strongly correlated and topological material microscopy:  simultaneous detection of magnetic and electric fields (down to the sub-single electron charge level); no invasive large magnetic fields or gradients; simultaneous micro- and macroscopic spatial resolution;  freedom from 1/$f$ flicker noise at low frequencies;  and the complete decoupling of probe and sample temperatures.  This latter feature is important since cooling topological insulator samples below $\sim$100 K is typically necessary to maximize sample resistivity.

\begin{figure}[t]
\includegraphics[width=0.475\textwidth]{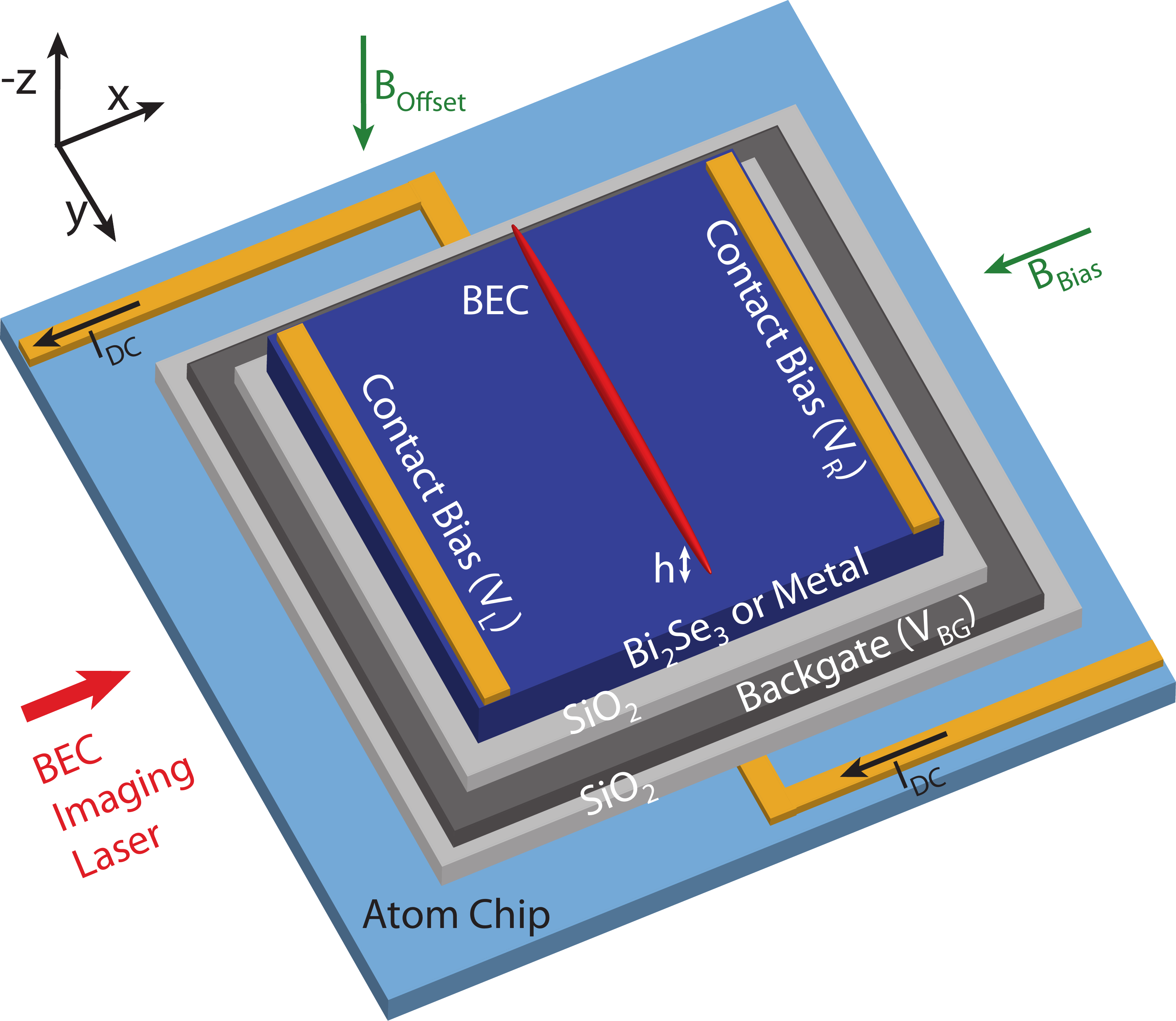}
\caption{(color online). Schematic of device under consideration.  Contacts induce a longitudinal current across the channel, which is back-gated to be within the bulk gap for the case of Bi$_2$Se$_3$.   The density of a BEC held in a cigar-shaped atom chip magnetic microtrap is distorted depending on the direction of current flow underneath.  The atom chip supports a Z-shaped gold wire that, with current $I_{DC}$ and homogeneous bias magnetic field $B_{\text{Bias}}$ along $-\hat{x}$, creates a magnetic trap above the substrates (see Sec.~\ref{ACM}).  A $\hat{z}$-oriented magnetic field $B_{\text{Offset}}$ allows the BEC to be shifted laterally in the $\hat{x}$ direction.  $h$ is the height of the BEC above the material. }
\label{fig:Schematic}
\end{figure}

We begin by describing atom chip microscopy~\cite{Schmiedmayer05_Nature,Schmiedmayer06_APL,Aigner:2008} and conclude with a scheme to measure the surface-to-bulk conductance ratio from the resulting DC magnetic field using this technique.  In support of this scheme, we calculate spatially resolved currents  to understand the effect of doping on surface transport in Bi$_2$Se$_3$ thin films.  Bi$_2$Se$_3$ is of particular interest because its bulk gap can be as high as 0.3 eV, though recent experiments have shown that the Fermi level in the bulk is usually pinned to the conduction band (CB) by Se vacancies, requiring either gating or doping to suppress bulk states~\cite{Ong10}. We focus our attention on the transport dynamics of the system depicted in Fig.~\ref{fig:Schematic}.  Contacts located on the top left and right edges of the system induce a longitudinal current across the thin film channel while the back gate tunes the chemical potential. We compare current profiles in undoped and doped Bi$_2$Se$_3$ thin films and to that of a conventional metal conductor.    We show that  a corrugated surface on a topological insulator, in which columns of material are removed, e.g., with a focused ion beam (FIB), to form trenches on the top surface, provides a unique environment to magnify the contrast between surface and bulk current and allows one to extract the surface-to-bulk conductance ratio from single-shot or multishot atom chip microscopy measurements.

\section{Imaging transport via atom chip microscopy}\label{ACM}

We now describe the atom chip trapping technique~\cite{Reichel01a,Schmiedmayer02,Zimmermann07}.  The proposed transport probe of TIs  consists of a collection of trapped ultracold or Bose-condensed neutral atoms.  While simple quadrupole magnetic traps formed from anti-Helmholtz coils---and variations thereof which produce harmonic traps---are sufficient for exploring many aspects of ultracold atomic physics, such macroscopic coils are poorly suited for accurately bringing gases within microns of a surface.
 
 Shrinking coils to the micron scale can greatly enhance the trap's gradient and curvature.  A microtrap's gradient and curvature scale as $I/r^2$ and $I/r^3$, respectively, where $I$ is the wire current and $r$ is the trap center-to-wire distance. Currents of $\sim$100 mA and wire cross sections of a few square microns are required to obtain high trap gradients ($\nabla B \propto B_\text{Bias}^2/I$) and curvatures ($\omega\propto \nabla B/\sqrt{B_\text{Offset}}$) at small $r$. Unfortunately, mounting such coils in a UHV chamber suitable for laser cooling is not practical.  Fortunately, fields with similar trap gradient and curvature scaling laws may be obtained from microfabricated wires on a planar substrate.  When combined with a weak, easily produced homogenous bias field $B_\text{Bias}$, such microwires create extremely tight magnetic traps for atoms suspended above the surface~\cite{Libbrecht95a}.  Full accessibility for nearby solid-state materials is preserved. The robust loading, confinement, and detection of ultracold atoms using chip-based traps has been demonstrated down to $h\approx1$ $\mu$m~\cite{Vladan04,Kruger04,Kruger05,Schmiedmayer11}, including atom chip trap-based BEC production (see Ref.~\cite{Zimmermann07} for review).  Condensate lifetimes above dielectrics have been measured to be greater than 1 s with minimal atom loss at distance $h$~\cite{Vladan04}.  The same trap stability is expected over the thin metals employed here, where the metal thickness is less than the skin depth and $h$~\cite{Henkel99,Henkel01}:  Surface state thickness in ideal TIs should be less than the 150-nm thickness of the metallic mirror coated on top of the sample (over a thin insulating intermediate layer), and non-ideal, doped TIs naturally have $10^6\times$ larger bulk resistivities (1~$\Omega$~cm) than, e.g., gold~\cite{Ando10}.  We will not  discuss here the well-established atom chip trapping procedures, imaging techniques, or considerations of surface Casimir-Polder potentials, but rather refer the reader to the extant literature~\cite{Reichel01a,Schmiedmayer02,Vladan04,Zimmermann07,Fromhold11}.

While BEC production using atom chips is now a mature technology, a major spoiler of device functionality has been the disturbance of the otherwise smooth trap potential from the very current-carrying conductors that form it.  The meandering current in wires due to scattering centers---and thus magnetic field inhomogeneities at the ppm level---cause fragmentation of the trapped BEC into ``sausage-link" like mini-BECs~\cite{Zimmermann07}. This fragmentation inhibits matter wave transport once the BEC chemical potential is less than local potential maxima.  

However, this surprisingly sensitive susceptibility of atom traps to magnetic field perturbations is a feature we propose to exploit for the study of transport in TIs.  By a simple rearrangement of the trapping bias field and microwire current path, atoms can be placed far away from the trapping wire ($>100$ $\mu$m) but within microns above a material whose magnetic and electric field inhomogeneities are of interest.  Figure~\ref{fig:ACM} depicts the operating principle of the atom chip microscope in which the density of a BEC is perturbed by a TI sample without adverse affects from the trapping wire itself. A recent experiment demonstrated the use of a Rb-based atom chip to image current flow in a room temperature gold wire with 10-$\mu$m resolution (3-$\mu$m magnetic field resolution) and sub-nT sensitivity~\cite{Schmiedmayer05_Nature,Schmiedmayer06_APL}.  Imaging at $h=3$ $\mu$m provided  3-$\mu$m-resolution of sub-mrad deviations in current flow angle~\cite{Aigner:2008}.  With improvements to  imaging systems, e.g., using high-numerical and aberration-corrected lens systems~\cite{Esslinger11}, resolution of transport flow at the 1 to 2-$\mu$m-level should be possible.

The atom chip microscopy measurements in this proposal requires the easily obtained confinement of a cigar-shaped BEC within $h = 2$--$10$ $\mu$m~\cite{Vladan04} from a TI surface. The axis of the BEC lies along $\hat{y}$ and can be positioned anywhere along $\hat{x}$ for imaging the magnetic field from the transport flow between the two bias contacts. With easily achievable microwire currents $I$ and bias magnetic fields $B_\text{Bias}$, the harmonic trapping potential may possess transverse frequencies $\omega_{\perp}$ approaching 1 kHz while maintaining the axial frequency below 10 Hz.  A field of $\sim$ 1 G at the trap minimum is achievable, which serves to prevent loss and heating from Majorana spin flips.  Small inhomogeneous fields $B_{\perp}$ transverse to the cigar-shaped BEC do not affect the density of the BEC due to the high transverse trapping frequencies.  However, inhomogeneous fields $B_{\parallel}$ along the BEC axis, even at the nT level, can easily perturb the BEC density due to its low chemical potential and weakly confining trap frequency.   Thus, the cigar-shaped, quasi-1D BEC serves as a vector-resolved magnetic field sensor, measuring field modulations along the condensate axis $\delta B_{\parallel}(y) = \mu\mu_{B}\hbar\omega_{\perp}\sqrt{1+4a_{s}n(y)}$, where $\mu=m_{F}g_{F}$ is the atomic magnetic moment, $a_{s}$ is the s-wave scattering length of the atoms, and $n(y)$ is the BEC's 1D density~\cite{Aigner:2008}.  The source current is derived from the local magnetic field map via  application of the Biot-Savart law~\cite{Schmiedmayer05_Nature,Schmiedmayer06_APL,Aigner:2008}.  Thus, the BEC serves to image transport as well as the local magnetic field inhomogeneties:  The sensed fields along $\hat{y}$ primarily arise from transport transverse to the BEC axis---i.e., in the $yz$-plane---allowing the BEC to detect deviations in both the depth and angle of transport.  Though for the measurements of surface--to--bulk conductivity ratio presented below in Sections~\ref{fieldfromtransport} and~\ref{StoBulk}, the condensate is primarily sensitive to modulation in $\hat{z}$ rather than angular deviation in $\hat{y}$.   This is due to the very broad sample (and trenches) in $\hat{y}$, resulting in a $\hat{y}$-field arising from a sheet, rather than line, current at different depths in $\hat{z}$ in the regions of imaging interest.  

Ultracold, but thermal, gases may be used as well for better dynamic field range, though at the price of lower field sensitivity compared to a BEC:  $\delta B_{\parallel}(y) = k_{B}T\log n(y)/\bar{n}$~\cite{Aigner:2008}.  The atomic gas density may be imaged above a surface assuming the top surface is made reflective with a $\sim$150 nm-thin metal film on thin insulator.  With high-resolution BEC imaging optics, the distance $h$ of the gas from the TI surface sets the transport imaging resolution of the atom chip microscope.  The Casimir-Polder potential and electrostatic patch field effects can distort magnetic traps and limit their lifetime within $\leq$1-$\mu$m from a surface~\cite{Vladan04,Cornell04,Cornell07,Cornell07b}.  However, ultracold gases have been confined and imaged in atom chip traps at $h\approx1$ $\mu$m and patch potentials have not been observed to affect such traps above at least 5~$\mu$m~\cite{Kruger04,Kruger05,Schmiedmayer11}.  Since the magnitude of measured fields scale with potential driving the current, all constant perturbations due to Casimir-Polder and patch fields can be calibrated out of the transport imaging measurement.   Moreover, measurements of the surface--to--bulk conductivity ratio, discussed in Sections~\ref{fieldfromtransport} and~\ref{StoBulk}, do not require resolution better than, e.g., 5 $\mu$m, because the feature sizes of the trenched TI can be on the tens of micron length scale.  Thus, only $h\geq5$ $\mu$m  surface heights are needed, and at these heights, Casimir-Polder and patch field effects will not significantly affect the trapping potential for atom chip microscopy even when imaging above trenched regions.

\section{Quantum transport calculation}\label{sec:QTC}

We perform quantum transport calculations on doped and undoped Bi$_2$Se$_3$ thin films using the non-equilibrium Green's function (NEGF) formalism~\cite{Datta00}.  These calculations allow us to examine current flow in topological insulators along smooth and corrugated surfaces and compare these transport profiles with that of a single orbital metal. 

For the TI, the 4-orbital effective Dirac Hamiltonian of Bi$_2$Se$_3$ is~\cite{Zhang:2009}:
\begin{equation}\label{eq:HkBiSe}
\begin{aligned}
{\cal H}_{TI}({\bf k}) = &\epsilon_0({\bf k})\mathbb{I} + {\cal M} ({\bf k}) \Gamma_0 \\
+ &(A_2k_x)\Gamma_1 + (A_2k_y)\Gamma_2 + (A_1k_z)\Gamma_3,
\end{aligned}
\end{equation}
where $\epsilon_0({\bf k})$ = C + D$_1$k$_z^2$ + D$_2$(k$_x^2$+k$_y^2$) and ${\cal M}({\bf k})$ = M - B$_1$k$_z^2$ - B$_2$(k$_x^2$ + k$_y^2$).  The system is in the basis $\Gamma_a = \tau^x \otimes \sigma^a$; $\Gamma_0 = \tau^z \otimes \mathbb{I}$, where $a= 1,2,3$.  By fitting the band structure to {\textit{ab initio}} calculations~\cite{Liu:2010}, the material parameters of Bi$_2$Se$_3$ that we use in this work are M = 0.28 eV, A$_1$ = 2.2~eV~$\AA$, A$_2$ = 4.1 eV $\AA$, B$_1$ = 10.0 eV $\AA^2$, B$_2$ = 56.6~eV~$\AA^2$, C = -0.0068~eV, D$_1$ = 1.3 eV $\AA^2$, and D$_2$ = 19.6 eV $\AA^2$. We discretize $\cal{H}({\bf k})$ into a nearest-neighbor real-space cubic lattice basis suitable for low energy transport (with a lattice constant $a_0=4$ $\AA$, see Appendix \ref{sec:Hr}), and we evaluate spatially resolved current from point $\vec{r}_1$ to $\vec{r}_2$ with
\begin{equation}\label{eq:SpatialCurrent}
\begin{aligned}
I(\vec{r}_1 \rightarrow \vec{r}_2) &= \frac{ie}{\hbar}\int \frac{dE}{2\pi} \\ [ & H(\vec{r}_{12})(G^n(\vec{r}_{21};E)-G^p(\vec{r}_{21};E)) \\
-&H(\vec{r}_{21})(G^n(\vec{r}_{12};E)-G^p(\vec{r}_{12};E))].
\end{aligned}
\end{equation}
$G^{n,p}(\vec{r}_{12};E)$ are the electron and hole correlation functions calculated within NEGF~\cite{SDattaQT} and $\vec{r}_{12}$ represents the off-diagonal block connecting sites $\vec{r}_1$ and $\vec{r}_2$, which is only nonzero for nearest neighbors.  

We define the Hamiltonian for a metal channel used for comparison with the TI systems with an isotropic single-orbital tight binding Hamiltonian,
\begin{equation}
{\cal H}_M = -t \sum_{\langle m,n \rangle} c_m^\dagger c_n + \text{H.c.},
\end{equation}
where the hopping energy is set to $t=3$ eV, $c_m^\dagger$ creates a fermion at site $m$ and $\langle m,n \rangle$ denotes nearest-neighbor bonds in all three spatial directions.  
The corresponding DC magnetic field from transport in either the TI or the metal is calculated from the current profile using the Biot-Savart law
\footnote{The magnetic susceptibility of Bi$_2$Se$_3$ is less than 10$^{-6}$ cm$^3$ mol$^{-1}$~\cite{Uemura:1977}.}.

\begin{figure}[t]
\includegraphics[width=0.475\textwidth]{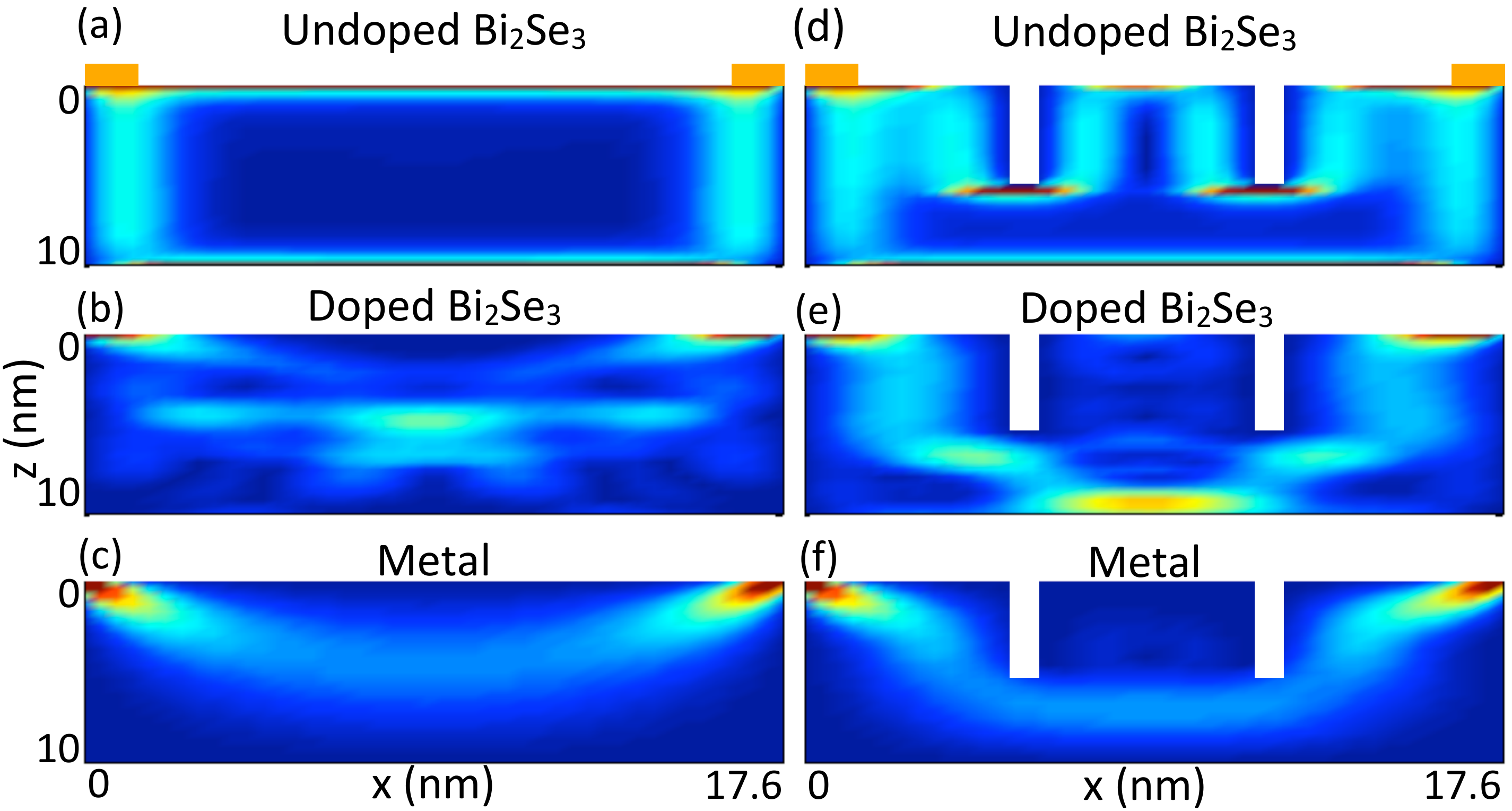}\caption{(color online).  Current profile, where red/bright (blue/dark) denote high (low) current density in: (a) and (d) undoped Bi$_2$Se$_3$; (b) and (e) doped Bi$_2$Se$_3$; and (c) and (f) metal thin films.  Positions of contact leads shown as gold rectangles in  (a) and (d); leads in the other panels not shown. (a) Current is closely tied to within 3 nm of the top and bottom surfaces of the undoped TI Bi$_2$Se$_3$.  (b) Doping by raising the chemical potential to 0.2~eV opens a parallel bulk conducting path as carriers reach the CB.  Backscattering in the CB marginally decreases total current by a few percent. The structure in the current density is due to finite simulation size effects.  (c) Carriers in the metal immediately diffuse and flow through the bulk of the channel.  The current profile of the metal is normalized so that total contact current is equivalent to the undoped Bi$_2$Se$_3$ profile.  (d--f) Current flow around two trenches in the top surface of the material.  Trenches in these plots are 1~nm wide, 5~nm deep, and separated by 7~nm.  (d) Current hugs the upper surface contour in  the undoped Bi$_2$Se$_3$ sample, but (e) fails to do so in the doped Bi$_2$Se$_3$.  (f) As in the doped TI, the trenches in a metal serve to spatially low-pass filter the $\hat{z}$ current modulation seen in the undoped TI.  Atom chip microscopy can image depth of the current flow between the trenches, thus providing a measure of surface--to--bulk conductivity.}
\label{fig:IdealSurfPlotxz}
\end{figure}

\subsection{Transport profiles}

We first describe numerical results comparing the spatially resolved current profiles for thin films of undoped Bi$_2$Se$_3$,  doped Bi$_2$Se$_3$, and metallic thin films in Fig.~\ref{fig:IdealSurfPlotxz}.  The dimensions of the system are limited by computational time to a channel 17.6 nm (44 sites) in $\hat{x}$ by 10.0~nm (25 sites) in $\hat{z},$  and with periodic boundary conditions in the transverse $\hat{y}$ direction, which is translationally invariant in our geometries. The contacts are placed only on the top surface ($z=0$) at $x=0$ and 17.6~nm and stretch along the entire $y$-direction.  The contact biases are set to V$_L$= -V$_R$ = 0.09~V so that injected carriers are confined within the bulk gap of $M = 0.28$~eV and thus pass only through surface states if the system is undoped.  Current flows along the topologically nontrivial surface states of the undoped Bi$_2$Se$_3$ [see Fig.~\ref{fig:IdealSurfPlotxz}(a)], decaying into the bulk with a length scale of $\sim$3~nm along the top and bottom surfaces and $\sim$5~nm along the side surfaces. The discrepancy is due to the anisotropy in the effective velocity in the $\hat{z}$ and $[\hat{x}, \hat{y}]$ directions, a consequence of the anisotropic quintuple layer structure of Bi$_2$Se$_3$.  We note that the current profile of the undoped system qualitatively retains this shape for all biases within the bulk gap.

Transport in an undoped, ideal Bi$_2$Se$_3$ sample with a perfectly insulating bulk is carried only through surface states. Since most Bi$_2$Se$_3$ thin films are not ideal, however, there is a finite bulk conductivity, and it is desirable to devise a scheme which can show clear evidence of the topological surface states in the presence of bulk doping. We show in the following section that our proposed geometry provides a means to measure the degree of doping in the system.

Recent experiments~\cite{Steinberg:2010} on Bi$_2$Se$_3$ thin films point to an inhomogeneous distribution of doping in which the chemical potential is smoothly raised along the bottom half of the TI.  However, we simply assume a homogenous doping profile of 0.2~eV above the Dirac point of the surface states. We have checked that our qualitative conclusions do not change if different  doping profiles are used. Figure~\ref{fig:IdealSurfPlotxz}(b) shows the resulting current profile when the chemical potential is raised to 0.2~eV above the Dirac point. A  parallel conducting path in the bulk limits topological flow along the top and bottom surface and there are significant  increases in the total current.

For completeness, we study  the transport behavior in a simple metallic thin film.  Such transport starkly differs from  ideal topologically nontrivial systems.  Hopping in the metal is isotropic, and we see in Fig.~\ref{fig:IdealSurfPlotxz}(c) that current diffuses into the gapless bulk.  (The current profile is normalized to have the same total current the same as in the undoped Bi$_2$Se$_3$ simulation.)  The metal retains this current profile regardless of bias strength or hopping energy. As a check on the validity of these numerical results, we calculate that current in the metal  flows around circular patches of (non-metallic) disorder at approximately $45^{\circ}$ angles (see Appendix~\ref{metallic}).  This is in good agreement with analytical calculations and with the angular dependence of current in a gold wire as experimentally observed using atom chip microscopy~\cite{Aigner:2008}.

\subsection{Transport around  corrugations}

Doped and undoped Bi$_2$Se$_3$ channels are capable of effecting different transport characteristics than topologically trivial materials, such as a simple metal.  However, as Fig.~\ref{fig:IdealSurfPlotxz}(b) shows, doping can mask TI-specific transport signatures.  We next seek to devise a channel geometry whose magnetic field response will provide a distinct signature of topological current flow.  We show in Figs.~\ref{fig:IdealSurfPlotxz}(d--f) that the current profiles in Figs.~\ref{fig:IdealSurfPlotxz}(a--c) change dramatically when two trenches 1-nm (2 sites) wide in $\hat{x}$ and 5-nm (12 sites) deep are formed on the top surface.  In the undoped TI, Fig.~\ref{fig:IdealSurfPlotxz}(d) shows that current flows around the trench and along the surface between the trenches.  In this configuration, with a 7-nm separation between trenches, we find that over 92\% of top surface current  flows within 3-nm of the top surface \emph{between} trenches, with only a small degradation due to bulk tunneling.  We note that the conductivity does not change when trenches are added, as the contacts still have access to the topologically protected surface states. 

 If the bulk doping is sufficient to open a parallel conducting path, however, surface and bulk states will hybridize and allow carriers to move  through the bulk instead of closely following the engineered geometric path, as shown in Fig.~\ref{fig:IdealSurfPlotxz}(e).   We find that top surface current between the trenches drops monotonically and approximately linearly as doping strength in the bulk is increased from zero to 0.2~eV. (See Fig.~\ref{fig:IsvsBiasBiSe} and Appendix~\ref{dependance} for the dependence of surface current on doping in the bulk.)  Current is pushed even further into the bulk for the simple metal [Fig.~\ref{fig:IdealSurfPlotxz}(f)], with otherwise little qualitative change. As expected, there is a decrease in conductivity of the trenched metal which is proportional to the change in effective cross-sectional area.

\section{Magnetic field from transport}\label{fieldfromtransport}

\begin{figure}[t]
\includegraphics[width=3.4in]{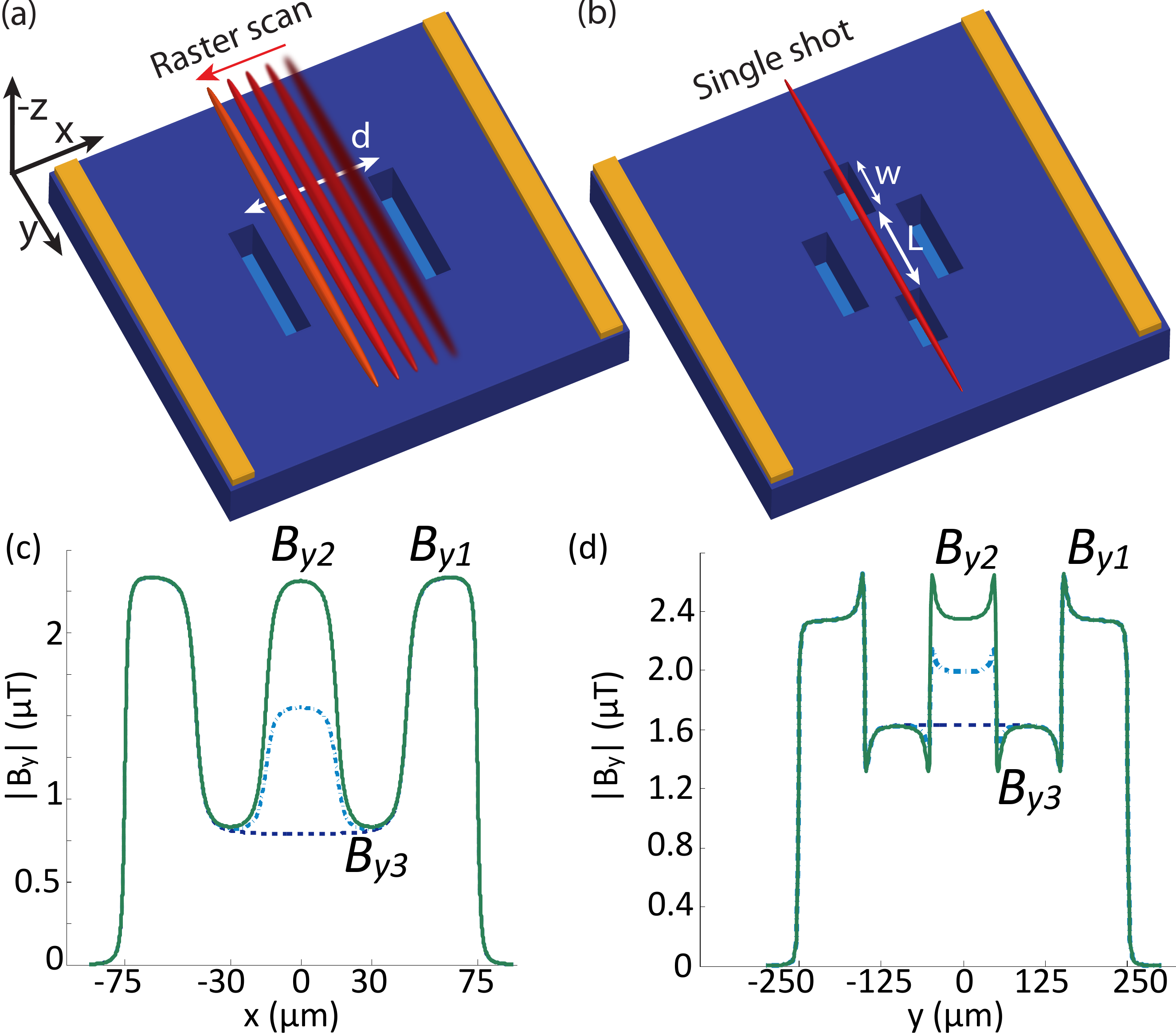}
\caption{(color online).   Current flow around trenches in the material differs depending on whether the material is an ideal TI, a metal, or a TI with non-zero bulk conductivity like Bi$_{2}$Se$_{3}$.   (a) Raster scan method for observing the field along $\hat{x}$.  BECs are sequentially created and imaged at positions along $\hat{x}$ to form the field profile in panel (c).    (b) Single-shot detection method in which a single BEC can image the field profile in panel (d).  (c)  Field $B_{y}$ from the raster scan method depicted in panel (a) is plotted 2~$\mu$m from surface.  The Bi$_2$Se$_3$  sample is 150~$\mu$m long and 10~$\mu$m wide.   The two trenches are 30~$\mu$m wide and 5~$\mu$m deep and are separated by $d = 30$~$\mu$m. (d) Field $B_{y}$ from the single-shot method depicted in panel (b) and plotted 2~$\mu$m from surface. The Bi$_2$Se$_3$  sample is 500~$\mu$m long and 500~$\mu$m wide.  The two trenches along $\hat{x}$ are 100~$\mu$m wide and 5~$\mu$m deep and are separated by 30~$\mu$m.  The two trenches along $\hat{y}$ are $w=100$~$\mu$m wide, $20$~$\mu$m long and separated by $L = 100$~$\mu$m.  Panels (c) and (d) show the magnetic field response for a system with surface--to--bulk current ratio of 100\% (green line), 50\% (dashed-dot light blue line), and 0\% (dashed dark blue line).  The magnetic fields are equivalent away from the midpoint between the trenches, while the magnetic field for the system with a 50\% or 0\% current ratio is less than for a 100\% ratio.  The magnetic field response follows this behavior for all current ratios, smoothly and monotonically decreasing as a function of doping level until the Fermi level reaches the conduction or valence bands.  }
\label{fig:Hcomparison2}
\end{figure}

\begin{figure}[t]
\includegraphics[width=2.5in]{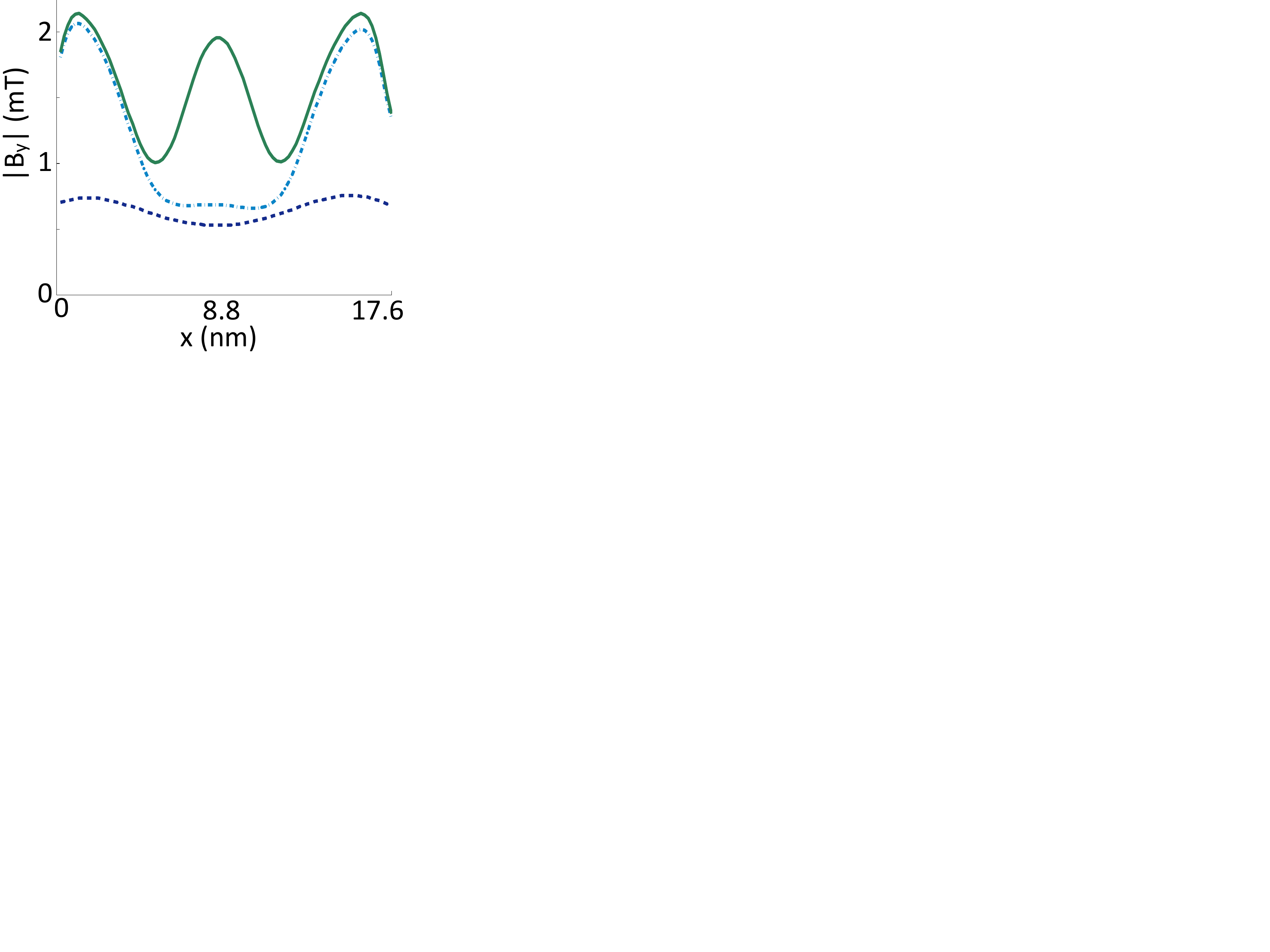}
\caption{(color online).  Transverse $\hat{y}$ component of the DC magnetic field $B_{y}$ produced by longitudinal transport of surface current density 5 $\mu$A/nm though the material with two trenches as in Fig.~\ref{fig:Schematic}.  Field is shown for undoped (green line) and doped (dashed-dot light blue line) Bi$_2$Se$_3$ and the metal model (dashed dark blue line).  The magnetic field is calculated $z=-1$~nm from the top surface, neglecting surface currents along the left and right sides of the system, and is plotted along the length of the material.  The surface current between the trenches in the undoped system creates a signature of topological current flow with an observable change in B$_y$ that is directly related to doping strength. Field magnitudes decrease in the doped system as total current is mitigated by conduction band backscattering.}
\label{fig:Hcomparison}
\end{figure}

Atom-chip microscopy does not measure the current directly, but rather the magnetic fields produced from the currents. The calculations depicted in  Fig.~\ref{fig:IdealSurfPlotxz} demonstrate a clear qualitative distinction between topologically non-trivial and trivial current flow in corrugated systems due to pinning of TI conducting state wavefunctions to the surface.  Such surface states, in contrast to bulk states, generate clear signatures in the magnetic field produced by DC current flow.  Corrugating the surface structure accentuates the surface current signature by inducing flow along paths with sharp discontinuities.  The field from such flow is fundamentally different than that of a metal or heavily doped TI.

Three methods may be used to distinguish an ideal TI with no doping [Fig.~\ref{fig:IdealSurfPlotxz}(d)] from one with doping [Fig.~\ref{fig:IdealSurfPlotxz}(e)].  Method (a) involves raster-scan imaging the field modulation along an $\hat{x}$-oriented line centered symmetrically above the trenches.  This field arises from the $\hat{z}$-modulation of the primarily $\hat{x}$-directed current [see Fig.~\ref{fig:Hcomparison2}(a)]. Method (b), similar to method (a), involves imaging the field modulation along a $\hat{y}$-oriented line centered symmetrically above the additional trenches depicted in Fig.~\ref{fig:Hcomparison2}(b).  Method (c) involves imaging the field arising from $\hat{y}$ current modulation around the corners of the trenches.   At a distance $z < -d$ from the surface, where $d$ is the trench separation in Fig.~\ref{fig:Hcomparison2}(a), the amplitude of the signal in method (c) is less than a factor of two different between a doped versus an undoped TI.  While current flow around the trench is in principle measurable, we  chose to focus this work on methods (a) and (b) because they provide a signal of much stronger contrast for distinguishing the degree of doping in a TI. 

In method (a), the $\hat{y}$-component of the magnetic field above a line connecting the centers of the parallel trenches is measured in a multishot, raster-scan fashion. Figure~\ref{fig:Hcomparison}(a) plots this $B_{y}$ magnetic field $z=1$ nm above the surface of each material system.  The undoped TI system shows a  peak in $B_y$ between the trenches due to surface current flow that approaches the maximum value halfway between the trenches.  The peak signature in the doped TI is reduced due to the spatial separation of bulk current; a small decrease in $B_y$ between the trenches is attributed to the surface current remnant.  The magnetic field response smoothly connects the undoped to the doped case as the surface-to-bulk conductance ratio increases.  This effect provides a  detection channel through which one may estimate the doping level from the surface-to-bulk conductance ratio, as discussed in more detail in Sec.~\ref{StoBulk}.  Additionally, the field from TIs show a much sharper increase to either side of the trench pair when compared with the conventional metal model: The topological system, regardless of doping, produces a  transport pattern distinct from simple metals because current does not immediately move into the bulk.

For an undoped TI, surface current magnitude does not appreciably change when trenches are added.  Surface conductance dominates, and $B_y$ is nearly as large between the trenches as it is outside of the trenches; the small discrepancy arises from current flowing in $\hat{z}$ around the trenches, which results in a $B_y$ reduction of roughly 5\% in our model system.  This effect would be less prominent in systems of larger size.  Elastic backscattering in the highly doped system via bulk channels, however, causes a slight decrease in the average magnitude of $B_y$.  

Unfortunately, our simulations are confined to small dimensions due to the computational cost of the formalism.  Matrix sizes of the 3D system quickly become intractable due to memory constraints even when translational invariance in the transverse direction is exploited. Thus, we are not able to numerically simulate systems that are of the necessary size for atom-chip microscopy. Fortunately, doping yields effects on transport regardless of trench size, channel dimensions, or aspect ratios, and the contrast in the  field from surface versus bulk current is insensitive to system size.  Over long length scales, inelastic scattering may harm the spin-orbit locking of the topologically protected surface state, but the existence of a conduction state at the surface remains regardless of system size.  In other words, while this scattering will dephase the spins in the system, we do not anticipate a significant depreciation in surface current flow as the impurity scattering would be too weak to induce significant bulk current flow or backscattering.  Therefore, the average magnitude of $B_y$ should remain nearly constant in large samples.

Current transport would generate a similar $B_{y}$ magnetic field if the system were expanded to be $10^{3}$ larger so that the channel and trench sizes in Fig.~\ref{fig:Hcomparison}(a) were of $\mu$m rather than nm scale.  Figure~\ref{fig:Hcomparison2}(c) shows $B_{y}$ at 1~$\mu$m from the surface when length scales are enlarged by $10^{3}$ to accommodate the atom chip microscope resolution.  For the same surface current density, the resulting field is $10^{3}$ smaller, $\sim$$10^{-6}$ T, which is well above the nT sensitivity of atom chip microscopy.  Currents $10^{3}$ smaller could be used to minimize resistive heating while still generating a detectable field signature.  The field calculation is described in further detail in Appendix~\ref{DCresponse}. Similarly measurable profiles at lower fields may be measured at $h\geq 2$ $\mu$m, which are heights more optimal for avoiding complications due to Casimir-Polder effects.   Adding more trenches to the TI surface in an array is expected to further accentuate the transport signals of interest.  

Method (b), while requiring four, rather than two, trenches, is a single-shot technique for measuring an analogous transport profile to that shown in Fig~\ref{fig:Hcomparison2}(c). (Additional complexity in fabricating multiple trenches is not prohibitive.)  Figure~\ref{fig:Hcomparison2}(b) shows how the ultracold atomic cloud---a length of 800~$\mu$m was imaged in Ref.~\cite{Kruger07} in one shot---could extend over two lateral trenches in addition to the space between the pair of trenches oriented along the $\hat{x}$ current flow.  The resulting $\hat{y}$ magnetic field  detected by the microscope, shown in Fig.~\ref{fig:Hcomparison2}(d), is qualitatively similar to that recorded in method (a).   Method (b) is much simpler and faster, however.

\section{Determination of surface--to--bulk conductance ratio}\label{StoBulk}

In order to determine the surface--to--bulk conductance ratio, we consider three points in the magnetic field plots of either Fig.~\ref{fig:Hcomparison2}(c) or (d): the maximum outside the trenches $B_{y1}$, the maximum directly between the trenches $B_{y2}$, and the minimum directly above one of the trenches $B_{y3}$.  The ratio of surface current $I_{s}$ between the trenches to surface current in an ideal system $I_{s,ideal}$ obeys the phenomenological equation:
\begin{equation}\label{eq:IsHyRelationship}
\frac{I_s}{I_{s,ideal}} = \frac{B_{y2}-B_{y3}}{B_{y1}-B_{y3}}.
\end{equation}
Equation \ref{eq:IsHyRelationship} yields a ratio of 92\% for the undoped system, compared to the actual value of 92\% in the simulation.  Current in $\hat{z}$ and limited system sizes cause a deviation in $B_y$ that is ameliorated by larger trench sizes and larger separation between the trenches.   Equation~\ref{eq:IsHyRelationship} provides a measure of $I_{s}$ given a microscopic model for $I_{s,ideal}$, and the surface--to--bulk conductivity ratio may be determined by measuring the total current $I$ through the system.  Fortunately, a microscopic model for $I_{s,ideal}$ is unnecessary if one calibrates the dependance of $I_{s}/I_{s,ideal}$ versus doping level, as we now explain.  

\begin{figure}[t]
\includegraphics[width=.485\textwidth]{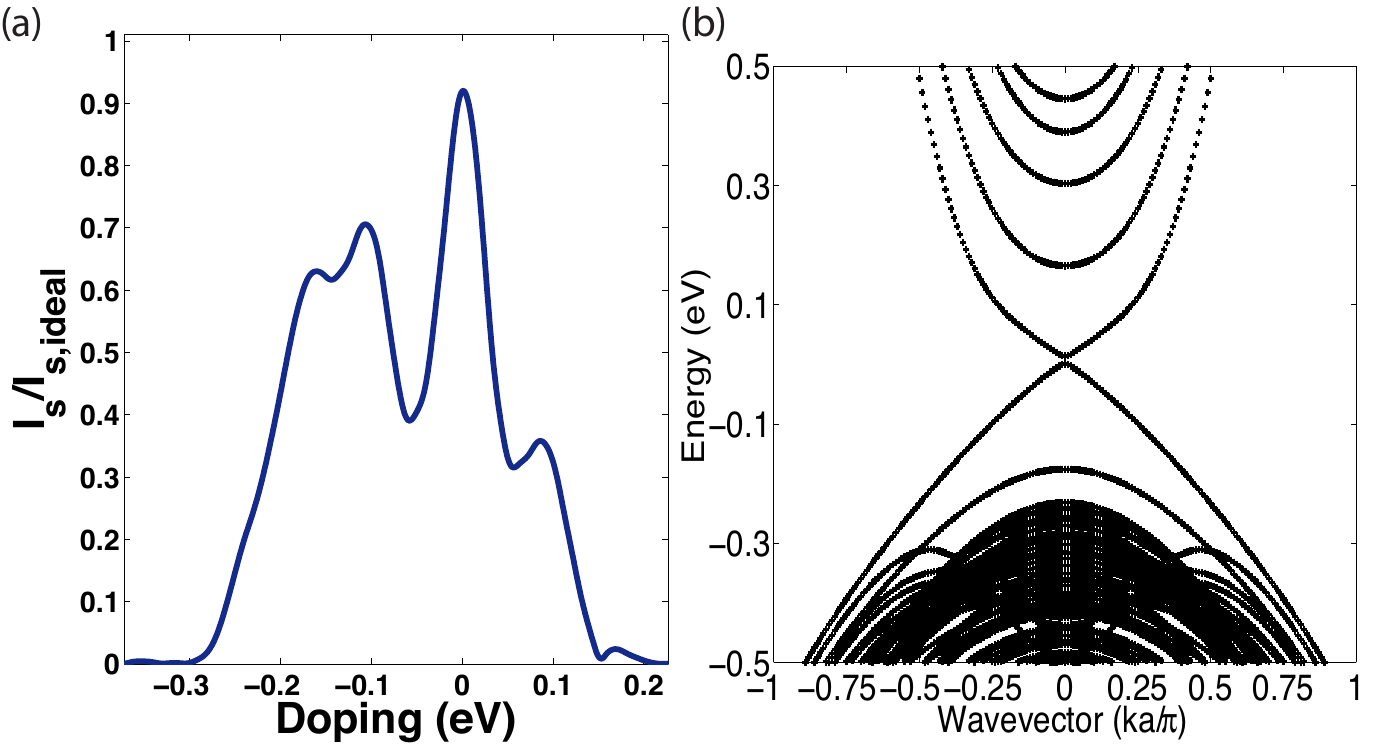}
\caption{(a) Plot of the surface current ratio between the trenches relative to the ideal surface current as a function of bulk doping strength for Bi$_2$Se$_3$. The highest energy carriers reach the first CB as soon as doping deviates from zero, causing a linear decay as more injected carriers have access to bulk states.  The highest energy carriers have access to the second CB at a doping strength $\agt|0.1|$~eV, and as a result, surface current exhibits.  (b) Plot of the Bi$_2$Se$_3$ band structure used for the simulation in panel (a).} 
\label{fig:IsvsBiasBiSe}
\end{figure}

Consider the decay of surface current between the trenches as doping strength varies:  Figure~\ref{fig:IsvsBiasBiSe}(a) shows how surface current between the trenches changes relative to the expected surface current in Bi$_2$Se$_3$.     When the backgate tunes the effective doping to be directly at the Dirac point of Bi$_2$Se$_3$ (compensating any intrinsic doping), surface current reaches 92\% of its ideal value when trenches are added. The relationship between surface current decay and doping strength is approximately linear until a bulk band is reached:  the additional bulk states cause surface current to modulate.  However, the overall effect near the Dirac point is monotonic and allows for the magnetic field response to smoothly map the transition from the undoped to the doped regimes.

The backgate in Fig.~\ref{fig:Schematic} allows one to tune the effective doping level regardless of intrinsic doping level.  By measuring $I_{s}/I_{s,ideal}$ in the multiple (a) or single shot (b) method of Section~\ref{fieldfromtransport}, one obtains a full $I_{s}/I_{s,ideal}$ curve as in Fig.~\ref{fig:IsvsBiasBiSe}(a). The curve in Fig.~\ref{fig:IsvsBiasBiSe}(a) is less monotonic than it would be had the simulation been immune to finite-size effects.  Experimental samples would be larger and exhibit fewer modulations.  Nevertheless, such a curve will always have its global maximum at the point where the Fermi level reaches the Dirac point.  This is also where $I_{ideal}\approx I_s$, and combined with a two-probe transport measurement of $I$ at high gate bias, such a measurement provides a model-independent calibration of the $I_{s}/I_{s,ideal}$ curve.  Mobility ratio in the bulk and surface can be determined by comparing the total current at this Dirac point backgate bias with a backgate bias far from this peak (e.g.,  near $\pm0.3$~eV in Fig.~\ref{fig:IsvsBiasBiSe}). Thus, determination of the bulk--to--surface conduction ratio can be accomplished in a relatively model-independent fashion.  (See Appendix~\ref{dependance} for discussion of the $I_{s}/I_{s,ideal}$ curve and band structure in the idealized TI case.)

\section{Conclusion}

Ultracold atom chip microscopy is capable of sensing a magnetic field signature of topological current flow.  Similarly, vector field imaging with diamond NV centers~\cite{Lukin08,Wrachtrup10} may be able to sense the field from transport.  However, the higher spatial resolution of NV centers may be counterbalanced by their comparatively low DC field sensitivity (which is maximal at kHz frequencies), lower field dynamic range, and the longer scan time needed to paint a picture of the transport flow.  (Though NV arrays with field detection alignment are under development.)  By contrast, the mm-long ultracold atomic gas provides transport images in a single shot with $10^{3}$ micron-sized pixels. 

We have shown that the contrast in magnetic field from transport in corrugated topological insulators provides a single-shot, relatively model-independent method for determining surface--to--bulk conductivity in Bi$_2$Se$_3$ thin films encumbered with Se vacancies.  Realization of BECs coupled to fields from TIs or topological--superconductor heterostructures may open avenues for quantum hybrid circuits involving atomic ensemble quantum memory and qubit transduction.

\begin{acknowledgments}
We would like to thank James Analytis, Ian Fisher, Lance Cooper, Peter Abbamonte, Richard Turner, and Matthew Naides for enlightening discussions.  We acknowledge generous support from the U.S. Department of Energy, Office of Basic Energy Sciences, Division of Materials Sciences and Engineering under awards \#DE-SC0001823 (B.L.L.) and  \#DE-FG02-07ER46453 (T.L.H.), and AFOSR award \#FA9550-10-1-0459 (B.D. and M.J.G.).
\end{acknowledgments}

\appendix
\begin{figure}[t]
\includegraphics[width=3.5in]{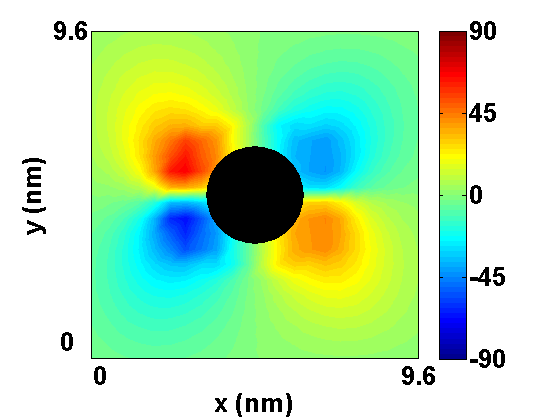}
\caption{(color online). Plot of angle dependence as electrons flow from left to right around a circular impurity in the metal model.  Carriers flow around the single impurity primarily at 45$^\circ$ angles, diffusing back to a uniform longitudinal current profile afterward.} 
\label{fig:MetalSingleImp}
\end{figure}

\section{Lattice Hamiltonian for Bi$_2$Se$_3$}\label{sec:Hr}
The $k$-space Dirac Hamiltonian for Bi$_2$Se$_3$ can be Fourier transformed into the real space Hamiltonian:
\bea\label{eq:Hrexpansion}
&{\cal H}_{TI} &= \sum_{lmn} c^{\dagger}_{lmn}\left [\left(C+\frac{D_1+2D_2}{a_0^2}\right)\mathbb{I}\right. \\
&+& \left. \left(M-\frac{B_1+2B_2}{a_0^2}\right)\Gamma_0\right]c^{\phantom{\dagger}}_{lmn}  \nonumber\\
&+&  c_{lmn}^\dagger  \left[\frac{B_2}{2a_0^2}\Gamma_0 - \frac{D_2}{2a_0^2}\mathbb{I} - \frac{iA_2}{2a_0}\Gamma_1\right]c^{\phantom{\dagger}}_{l+1mn} + \text{H.c.} \nonumber \\
&+&  c_{lmn}^\dagger  \left[\frac{B_2}{2a_0^2}\Gamma_0 - \frac{D_2}{2a_0^2}\mathbb{I} - \frac{iA_2}{2a_0}\Gamma_2\right] c^{\phantom{\dagger}}_{lm+1n}+ \text{H.c.} \nonumber \\
&+&  c_{lmn}^\dagger  \left[\frac{B_1}{2a_0^2}\Gamma_0 - \frac{D_1}{2a_0^2}\mathbb{I} - \frac{iA_1}{2a_0}\Gamma_3\right]c^{\phantom{\dagger}}_{lmn+1} + \text{H.c.}\nonumber,
\eea
where $c_{lmn}=\left[c_{A\uparrow,lmn}\;\; c_{A\downarrow,lmn}\;\; c_{B\uparrow,lmn}\;\; c_{B\downarrow,lmn}\right]^{T}$ ($c^{\dagger}_{lmn}$) destroys (creates) a fermion at site ($x = l$, $y = m$, $z = n$), $a_0$ is the lattice constant and the $\Gamma_{i}$ matrices and material parameters are defined in Section \ref{sec:QTC}.  Band structure calculations show the Dirac point occurs at $\sim$0.21~eV, with conduction band (CB) minimum and valence band (VB) maximum occurring at $\sim$0.31~eV and $\sim$-0.02~eV, respectively.  The back gate of the system tunes the Fermi energy to the Dirac point, where the real space Hamiltonian suitably describes low energy transport.  When the effective lattice constant is set to 4~$\AA$, nearest neighbor hopping energies are large enough for on-site energies within $\pm$0.3 eV of the Dirac point.  The CB minimum is just over 0.1~eV above the Dirac point, and the bias configuration is accordingly set to V$_L$$ = -$V$_R$ = 0.09~V so that all injected carriers have energies within the bulk gap.

\section{Metallic transport around  defects}\label{metallic}
The first use of atom chip microscopy demonstrated that current tends to move around impurities in a metal at 45$^\circ$ angles, as expected from analytical calculations~\cite{Aigner:2008}. We replicate this observation as a cross-check of our numerics by placing a circular impurity 3~nm in diameter in the middle of the metal channel with bias configuration V$_L$ = -0.09~V and V$_R$ = 0.09~V. (Electron majority carriers flow from left to right.)  Figure~\ref{fig:MetalSingleImp} shows that electrons primarily flow around the circular impurity at 45$^\circ$ angles before returning to a uniform current flow due to dissipation.  The angular dependence is less prominent beyond the impurity, as dissipation is the cause of angular distortion in the metal rather than a sharp potential boundary as for the TIs. The single-orbital tight-binding Hamiltonian is thus a simple model that nevertheless produces a current profile retaining the salient physics of transport in disordered metals.  As is the case in the trenched system, the angular dependence retains this characteristic shape of the current profile regardless of bias range or system size.

\section{Effect of doping on surface current in idealized TI}\label{dependance}

\begin{figure}[t]
\includegraphics[width=.485\textwidth]{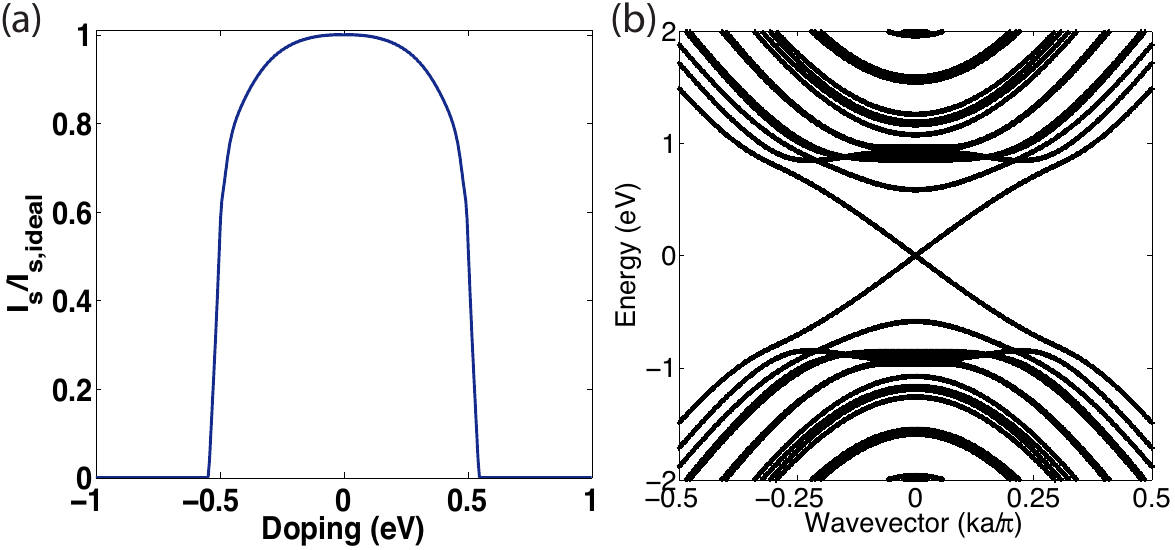}
\caption{(a) Plot of the ratio of surface current between the trenches relative to an idealized surface current as a function of bulk doping strength for an idealized TI.  The highest energy injected carriers begin to reach the CB minimum at doping strengths around $\pm0.5$~eV.
  (b) Plot of the  idealized TI  band structure used for the simulation in panel (a).} 
\label{fig:idealCalCurve}
\end{figure}

We perform an additional calculation with idealized TI parameters for the lattice Hamiltonian in Eq.~\ref{eq:Hrexpansion}: M = 1.0~eV, A$_1$ = A$_2$ = 1.0~eV~$\AA$, B$_1$ = B$_2$ = 1.0~eV~$\AA^2$, C = 0.0 eV, D$_1$ = D$_2$ = 0.0~eV~$\AA^2$ and $a_0=1$~$\AA.$ Figure~\ref{fig:idealCalCurve}(a) explores the surface current ratio with this toy model using the same bias configuration as in the Bi$_2$Se$_3$ case (V$_L$ = -V$_R$ = 0.09~V).  The gapless states are much more closely tied to the surface in this model, so tunneling between trenches is negligible when doping is zero, see Fig.~\ref{fig:idealCalCurve}. The ratio remains near unity until the CB minimum is reached, as the model has a much larger bulk gap.  However, there is a decline in the ratio once the first CB minimum is reached at $\pm0.5$~eV. 

\begin{figure}[t]
\includegraphics[width=.492\textwidth]{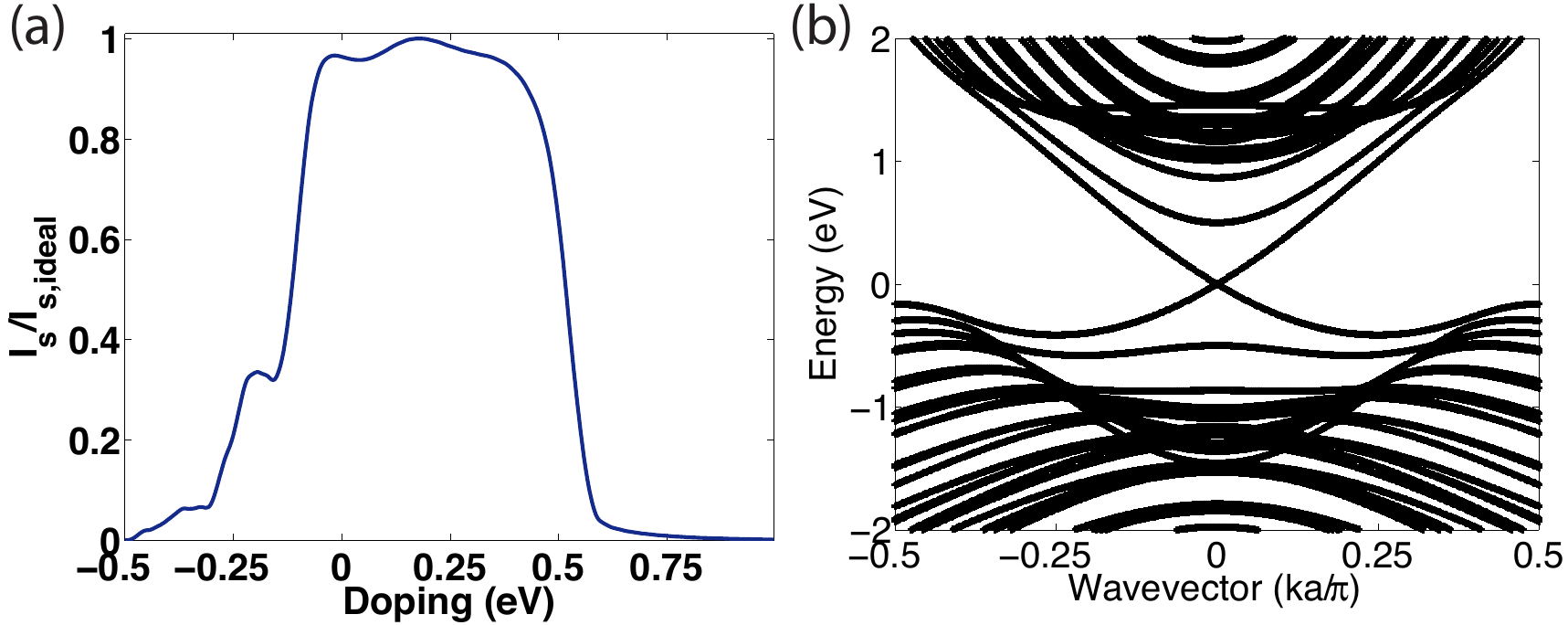}
\caption{Same as Fig.~\ref{fig:idealCalCurve}, but with particle-hole asymmetry.} 
\label{fig:idealCalCurveshifted}
\end{figure}

Particle-hole symmetry breaking biases the measurement of the backgate voltage at the $I_{s,ideal}=I_s$ point away from the Dirac point.  Such asymmetry is induced by setting $D_{1}=0.5$~eV~$\AA^2$.  Figure~\ref{fig:idealCalCurveshifted} shows simulations with particle-hole asymmetry demonstrating this effect.  The peak is offset from the Dirac point, but by an amount given by the changed dispersion (Fermi velocity) at the Dirac point. Once measured, this effect may be accounted for in the $I_s/I_{s,ideal}$ calibration.

\section{Analytic calculation of  magnetic field}\label{DCresponse}
The small systems sizes to which this formalism is limited prevents the trenches from completely isolating each component of the magnetic field response necessary for determining the surface--to--bulk conductance ratio.  Undesirable contributions from current around the trenches cause a significant deviation of roughly 5\% to the magnetic field response.  To show how this is mitigated with larger system sizes---and to produce the plots in Fig.~\ref{fig:Hcomparison}(c) and (d)---we compare results of the simulation to the analytical magnetic field response of a current carrying plate with equivalent geometry.  Using the Biot-Savart law, the magnetic field response for a current-carrying wire going from point x$_l$ to x$_r$ in the longitudinal direction is
\bea\label{eq:Hwire}
\vec{B}(\vec{r}) &=& \frac{\mu_0}{4\pi} \int_{x_l}^{x_r} \frac{I d\vec{l} \times \vec{r}}{r^3} \\
&=&  \frac{\mu_0Id}{4\pi} \int_{x_l}^{x_r} \frac{(\hat{z}\sin\theta - \hat{y}\cos\theta)dx'}{((x-x')^2+y^2+z^2)^{3/2}}\\ 
&=&\frac{\mu_0 I}{4 \pi d}(\hat{z}\sin\theta - \hat{y}\cos\theta) \\
&&\left[ \frac{x_r -x}{\sqrt{d^2+(x-x_r)^2}} - \frac{x_l-x}{\sqrt{d^2+(x-x_l)^2}} \right], \nonumber 
\eea
where $d=\sqrt{y^2+z^2}$ and $\theta = \tan^{-1}(y/z)$.  We assume the points at which the magnetic field will be measured lie along the $\hat{x}$ direction, directly above the middle of the channel, where y=0 and $\theta=0$.  The resulting magnetic field will have negligible $B_x$ and $B_z$ components due to the symmetry of a sheet current.  

We now seek to take into account the width of the current carrying surface.  According to Ampere's law, the total magnetic field a distance $d$ above a horizontal current sheet of finite width and infinite length is 
\bea\label{eq:Hwire}
\vec{B}(\vec{r}) &=& -\hat{y}\frac{\mu_0 j_0 d}{\pi} \int_{0}^{W/2} \frac{dy}{y^2+d^2} \\
&=&  -\hat{y}\frac{\mu_0 j_0}{\pi}\arctan\frac{W}{2d}, \nonumber 
\eea
where W is the width of the channel.  The magnetic field above a current plate thus decays proportionally to $\arctan(W/2d)$ rather than a simple $1/d$ dependence.  The change in the magnetic field magnitude caused by current moving in the bulk compared to current along the surface must be appreciable, thus the width should remain as small as possible so that the contrast remains high.   For example, a width of 10~$\mu$m is thin enough for the magnetic field to drop more than 50\% when the trenches are 5~$\mu$m deep and the atoms lie 1-2~$\mu$m from the surface.  

We  define two functions for the $\hat{y}$ magnetic field response due to $\hat{x}$ and $\hat{z}$-directed current, which are respectively
\bea\label{eq:HyIx}
&B&_{Ix}(x, y; x_l, x_r, z_0, j_m) \equiv  -\frac{\mu_0 j_m a_0}{4 \pi}\frac{z_0}{y^2+z_0^2} \\
&& \left[ \frac{x_r -x}{\sqrt{(x-x_r)^2+y^2+z_0^2}} - \frac{x_l-x}{\sqrt{(x-x_l)^2+y^2+z_0^2}} \right],\nonumber
\eea

\bea\label{eq:HyIx}
&B&_{Iz}(x, y; x_0, z_l, z_r, j_m) \equiv \frac{\mu_0 j_m a_0}{4 \pi} \frac{x}{x^2+y^2}\\
&&\left[ \frac{z_r}{\sqrt{(x-x_0)^2+y^2+z_r^2}} - \frac{z_l}{\sqrt{(x-x_0)^2+y^2+z_l^2}}  \right]. \nonumber
\eea
These functions are used to calculate the magnetic field response of a current configuration analogous to those seen in Fig.~\ref{fig:Hcomparison} but for larger dimensions.  We calculate the magnetic field response by summing over the entire width of the system.

 The simulations yield a surface current density of 4.9~$\mu$A nm$^{-1}$, in good agreement with the analytic solution for current in Bi$_2$Se$_3$,  
\begin{equation}
j_0=\frac{e^2}{h}k_FV_{app}=5.1 \,\mu A\, \text{nm}^{-1},
\end{equation}
where k$_F$ is the Fermi wave vector, for an applied bias V$_{app}$ = V$_L$ - V$_R$ = 0.18~V. We ignore side surface current and assume the channel is deep enough that surface current contributions to the magnetic field around the bottom corners may be neglected.  Homogeneous additions to the magnetic field will not change the calculation, as the surface--to--bulk ratio is determined by the magnetic field magnitude in between the trenches relative to the magnetic field above each trench.  A similar calculation was used to create the field profile in Fig.~\ref{fig:Hcomparison2}(d).


\begin{thebibliography}{56}%
\makeatletter
\providecommand \@ifxundefined [1]{%
 \@ifx{#1\undefined}
}%
\providecommand \@ifnum [1]{%
 \ifnum #1\expandafter \@firstoftwo
 \else \expandafter \@secondoftwo
 \fi
}%
\providecommand \@ifx [1]{%
 \ifx #1\expandafter \@firstoftwo
 \else \expandafter \@secondoftwo
 \fi
}%
\providecommand \natexlab [1]{#1}%
\providecommand \enquote  [1]{``#1''}%
\providecommand \bibnamefont  [1]{#1}%
\providecommand \bibfnamefont [1]{#1}%
\providecommand \citenamefont [1]{#1}%
\providecommand \href@noop [0]{\@secondoftwo}%
\providecommand \href [0]{\begingroup \@sanitize@url \@href}%
\providecommand \@href[1]{\@@startlink{#1}\@@href}%
\providecommand \@@href[1]{\endgroup#1\@@endlink}%
\providecommand \@sanitize@url [0]{\catcode `\\12\catcode `\$12\catcode
  `\&12\catcode `\#12\catcode `\^12\catcode `\_12\catcode `\%12\relax}%
\providecommand \@@startlink[1]{}%
\providecommand \@@endlink[0]{}%
\providecommand \url  [0]{\begingroup\@sanitize@url \@url }%
\providecommand \@url [1]{\endgroup\@href {#1}{\urlprefix }}%
\providecommand \urlprefix  [0]{URL }%
\providecommand \Eprint [0]{\href }%
\providecommand \doibase [0]{http://dx.doi.org/}%
\providecommand \selectlanguage [0]{\@gobble}%
\providecommand \bibinfo  [0]{\@secondoftwo}%
\providecommand \bibfield  [0]{\@secondoftwo}%
\providecommand \translation [1]{[#1]}%
\providecommand \BibitemOpen [0]{}%
\providecommand \bibitemStop [0]{}%
\providecommand \bibitemNoStop [0]{.\EOS\space}%
\providecommand \EOS [0]{\spacefactor3000\relax}%
\providecommand \BibitemShut  [1]{\csname bibitem#1\endcsname}%
\let\auto@bib@innerbib\@empty
\bibitem [{\citenamefont {Hasan}\ and\ \citenamefont {Kane}(2010)}]{Hasan10}%
  \BibitemOpen
  \bibfield  {author} {\bibinfo {author} {\bibfnamefont {M.~Z.}\ \bibnamefont
  {Hasan}}\ and\ \bibinfo {author} {\bibfnamefont {C.~L.}\ \bibnamefont
  {Kane}},\ }\href {\doibase 10.1103/RevModPhys.82.3045} {\bibfield  {journal}
  {\bibinfo  {journal} {Rev. Mod. Phys.}\ }\textbf {\bibinfo {volume} {82}},\
  \bibinfo {pages} {3045} (\bibinfo {year} {2010})}\BibitemShut {NoStop}%
\bibitem [{\citenamefont {Qi}\ and\ \citenamefont {Zhang}(2011)}]{Qi11}%
  \BibitemOpen
  \bibfield  {author} {\bibinfo {author} {\bibfnamefont {X.-L.}\ \bibnamefont
  {Qi}}\ and\ \bibinfo {author} {\bibfnamefont {S.-C.}\ \bibnamefont {Zhang}},\
  }\href {\doibase 10.1103/RevModPhys.83.1057} {\bibfield  {journal} {\bibinfo
  {journal} {Rev. Mod. Phys.}\ }\textbf {\bibinfo {volume} {83}},\ \bibinfo
  {pages} {1057} (\bibinfo {year} {2011})}\BibitemShut {NoStop}%
\bibitem [{\citenamefont {Fu}\ \emph {et~al.}(2007)\citenamefont {Fu},
  \citenamefont {Kane},\ and\ \citenamefont {Mele}}]{Fu:2007}%
  \BibitemOpen
  \bibfield  {author} {\bibinfo {author} {\bibfnamefont {L.}~\bibnamefont
  {Fu}}, \bibinfo {author} {\bibfnamefont {C.~L.}\ \bibnamefont {Kane}}, \ and\
  \bibinfo {author} {\bibfnamefont {E.~J.}\ \bibnamefont {Mele}},\ }\href
  {\doibase 10.1103/PhysRevLett.98.106803} {\bibfield  {journal} {\bibinfo
  {journal} {Phys. Rev. Lett.}\ }\textbf {\bibinfo {volume} {98}},\ \bibinfo
  {pages} {106803} (\bibinfo {year} {2007})}\BibitemShut {NoStop}%
\bibitem [{\citenamefont {Moore}\ and\ \citenamefont
  {Balents}(2007)}]{Moore:2007}%
  \BibitemOpen
  \bibfield  {author} {\bibinfo {author} {\bibfnamefont {J.~E.}\ \bibnamefont
  {Moore}}\ and\ \bibinfo {author} {\bibfnamefont {L.}~\bibnamefont
  {Balents}},\ }\href {\doibase 10.1103/PhysRevB.75.121306} {\bibfield
  {journal} {\bibinfo  {journal} {Phys. Rev. B}\ }\textbf {\bibinfo {volume}
  {75}},\ \bibinfo {pages} {121306} (\bibinfo {year} {2007})}\BibitemShut
  {NoStop}%
\bibitem [{\citenamefont {Qi}\ \emph {et~al.}(2008)\citenamefont {Qi},
  \citenamefont {Hughes},\ and\ \citenamefont {Zhang}}]{Qi:2008}%
  \BibitemOpen
  \bibfield  {author} {\bibinfo {author} {\bibfnamefont {X.-L.}\ \bibnamefont
  {Qi}}, \bibinfo {author} {\bibfnamefont {T.~L.}\ \bibnamefont {Hughes}}, \
  and\ \bibinfo {author} {\bibfnamefont {S.-C.}\ \bibnamefont {Zhang}},\ }\href
  {\doibase 10.1103/PhysRevB.78.195424} {\bibfield  {journal} {\bibinfo
  {journal} {Phys. Rev. B}\ }\textbf {\bibinfo {volume} {78}},\ \bibinfo
  {pages} {195424} (\bibinfo {year} {2008})}\BibitemShut {NoStop}%
\bibitem [{\citenamefont {Schnyder}\ \emph {et~al.}(2008)\citenamefont
  {Schnyder}, \citenamefont {Ryu}, \citenamefont {Furusaki},\ and\
  \citenamefont {Ludwig}}]{Schnyder:2008}%
  \BibitemOpen
  \bibfield  {author} {\bibinfo {author} {\bibfnamefont {A.~P.}\ \bibnamefont
  {Schnyder}}, \bibinfo {author} {\bibfnamefont {S.}~\bibnamefont {Ryu}},
  \bibinfo {author} {\bibfnamefont {A.}~\bibnamefont {Furusaki}}, \ and\
  \bibinfo {author} {\bibfnamefont {A.~W.~W.}\ \bibnamefont {Ludwig}},\ }\href
  {\doibase 10.1103/PhysRevB.78.195125} {\bibfield  {journal} {\bibinfo
  {journal} {Phys. Rev. B}\ }\textbf {\bibinfo {volume} {78}},\ \bibinfo
  {pages} {195125} (\bibinfo {year} {2008})}\BibitemShut {NoStop}%
\bibitem [{\citenamefont {Leijnse}\ and\ \citenamefont
  {Flensberg}(2011)}]{Flensberg11}%
  \BibitemOpen
  \bibfield  {author} {\bibinfo {author} {\bibfnamefont {M.}~\bibnamefont
  {Leijnse}}\ and\ \bibinfo {author} {\bibfnamefont {K.}~\bibnamefont
  {Flensberg}},\ }\href {\doibase 10.1103/PhysRevLett.107.210502} {\bibfield
  {journal} {\bibinfo  {journal} {Phys. Rev. Lett.}\ }\textbf {\bibinfo
  {volume} {107}},\ \bibinfo {pages} {210502} (\bibinfo {year}
  {2011})}\BibitemShut {NoStop}%
\bibitem [{\citenamefont {Ren}\ \emph {et~al.}(2010)\citenamefont {Ren},
  \citenamefont {Taskin}, \citenamefont {Sasaki}, \citenamefont {Segawa},\ and\
  \citenamefont {Ando}}]{Ando10}%
  \BibitemOpen
  \bibfield  {author} {\bibinfo {author} {\bibfnamefont {Z.}~\bibnamefont
  {Ren}}, \bibinfo {author} {\bibfnamefont {A.~A.}\ \bibnamefont {Taskin}},
  \bibinfo {author} {\bibfnamefont {S.}~\bibnamefont {Sasaki}}, \bibinfo
  {author} {\bibfnamefont {K.}~\bibnamefont {Segawa}}, \ and\ \bibinfo {author}
  {\bibfnamefont {Y.}~\bibnamefont {Ando}},\ }\href@noop {} {\bibfield
  {journal} {\bibinfo  {journal} {Phys. Rev. B}\ }\textbf {\bibinfo {volume}
  {82}},\ \bibinfo {pages} {241306} (\bibinfo {year} {2010})}\BibitemShut
  {NoStop}%
\bibitem [{\citenamefont {Xiong}\ \emph {et~al.}(2012)\citenamefont {Xiong},
  \citenamefont {Petersen}, \citenamefont {Qu}, \citenamefont {Hor},
  \citenamefont {Cava},\ and\ \citenamefont {Ong}}]{Ong11}%
  \BibitemOpen
  \bibfield  {author} {\bibinfo {author} {\bibfnamefont {J.}~\bibnamefont
  {Xiong}}, \bibinfo {author} {\bibfnamefont {A.}~\bibnamefont {Petersen}},
  \bibinfo {author} {\bibfnamefont {D.}~\bibnamefont {Qu}}, \bibinfo {author}
  {\bibfnamefont {Y.}~\bibnamefont {Hor}}, \bibinfo {author} {\bibfnamefont
  {R.}~\bibnamefont {Cava}}, \ and\ \bibinfo {author} {\bibfnamefont
  {N.}~\bibnamefont {Ong}},\ }\href@noop {} {\bibfield  {journal} {\bibinfo
  {journal} {Physica E}\ }\textbf {\bibinfo {volume} {44}} (\bibinfo {year}
  {2012})}\BibitemShut {NoStop}%
\bibitem [{\citenamefont {Analytis}\ \emph
  {et~al.}(2010{\natexlab{a}})\citenamefont {Analytis}, \citenamefont
  {McDonald}, \citenamefont {Riggs}, \citenamefont {Chu}, \citenamefont
  {Boebinger},\ and\ \citenamefont {Fisher}}]{Fisher10}%
  \BibitemOpen
  \bibfield  {author} {\bibinfo {author} {\bibfnamefont {J.~G.}\ \bibnamefont
  {Analytis}}, \bibinfo {author} {\bibfnamefont {R.~D.}\ \bibnamefont
  {McDonald}}, \bibinfo {author} {\bibfnamefont {S.~C.}\ \bibnamefont {Riggs}},
  \bibinfo {author} {\bibfnamefont {J.-H.}\ \bibnamefont {Chu}}, \bibinfo
  {author} {\bibfnamefont {G.~S.}\ \bibnamefont {Boebinger}}, \ and\ \bibinfo
  {author} {\bibfnamefont {I.~R.}\ \bibnamefont {Fisher}},\ }\href@noop {}
  {\bibfield  {journal} {\bibinfo  {journal} {Nature Physics}\ }\textbf
  {\bibinfo {volume} {6}},\ \bibinfo {pages} {960} (\bibinfo {year}
  {2010}{\natexlab{a}})}\BibitemShut {NoStop}%
\bibitem [{\citenamefont {Steinberg}\ \emph {et~al.}(2011)\citenamefont
  {Steinberg}, \citenamefont {Lalo\"e}, \citenamefont {Fatemi}, \citenamefont
  {Moodera},\ and\ \citenamefont {Jarillo-Herrero}}]{JarilloHerrero11}%
  \BibitemOpen
  \bibfield  {author} {\bibinfo {author} {\bibfnamefont {H.}~\bibnamefont
  {Steinberg}}, \bibinfo {author} {\bibfnamefont {J.-B.}\ \bibnamefont
  {Lalo\"e}}, \bibinfo {author} {\bibfnamefont {V.}~\bibnamefont {Fatemi}},
  \bibinfo {author} {\bibfnamefont {J.~S.}\ \bibnamefont {Moodera}}, \ and\
  \bibinfo {author} {\bibfnamefont {P.}~\bibnamefont {Jarillo-Herrero}},\
  }\href {\doibase 10.1103/PhysRevB.84.233101} {\bibfield  {journal} {\bibinfo
  {journal} {Phys. Rev. B}\ }\textbf {\bibinfo {volume} {84}},\ \bibinfo
  {pages} {233101} (\bibinfo {year} {2011})}\BibitemShut {NoStop}%
\bibitem [{\citenamefont {Chen}\ \emph {et~al.}(2009)\citenamefont {Chen},
  \citenamefont {Analytis}, \citenamefont {Chu}, \citenamefont {Liu},
  \citenamefont {Mo}, \citenamefont {Qi}, \citenamefont {Zhang}, \citenamefont
  {Lu}, \citenamefont {Dai}, \citenamefont {Fang}, \citenamefont {Zhang},
  \citenamefont {Fisher}, \citenamefont {Hussain},\ and\ \citenamefont
  {Shen}}]{Chen:2009}%
  \BibitemOpen
  \bibfield  {author} {\bibinfo {author} {\bibfnamefont {Y.~L.}\ \bibnamefont
  {Chen}}, \bibinfo {author} {\bibfnamefont {J.~G.}\ \bibnamefont {Analytis}},
  \bibinfo {author} {\bibfnamefont {J.-H.}\ \bibnamefont {Chu}}, \bibinfo
  {author} {\bibfnamefont {Z.~K.}\ \bibnamefont {Liu}}, \bibinfo {author}
  {\bibfnamefont {S.-K.}\ \bibnamefont {Mo}}, \bibinfo {author} {\bibfnamefont
  {X.~L.}\ \bibnamefont {Qi}}, \bibinfo {author} {\bibfnamefont {H.~J.}\
  \bibnamefont {Zhang}}, \bibinfo {author} {\bibfnamefont {D.~H.}\ \bibnamefont
  {Lu}}, \bibinfo {author} {\bibfnamefont {X.}~\bibnamefont {Dai}}, \bibinfo
  {author} {\bibfnamefont {Z.}~\bibnamefont {Fang}}, \bibinfo {author}
  {\bibfnamefont {S.~C.}\ \bibnamefont {Zhang}}, \bibinfo {author}
  {\bibfnamefont {I.~R.}\ \bibnamefont {Fisher}}, \bibinfo {author}
  {\bibfnamefont {Z.}~\bibnamefont {Hussain}}, \ and\ \bibinfo {author}
  {\bibfnamefont {Z.-X.}\ \bibnamefont {Shen}},\ }\href {\doibase
  10.1126/science.1173034} {\bibfield  {journal} {\bibinfo  {journal}
  {Science}\ }\textbf {\bibinfo {volume} {325}},\ \bibinfo {pages} {178}
  (\bibinfo {year} {2009})}\BibitemShut {NoStop}%
\bibitem [{\citenamefont {Hsieh}\ \emph {et~al.}(2009)\citenamefont {Hsieh},
  \citenamefont {Xia}, \citenamefont {Qian}, \citenamefont {Wray},
  \citenamefont {Dil}, \citenamefont {Meier}, \citenamefont {Osterwalder},
  \citenamefont {Patthey}, \citenamefont {Checkelsky}, \citenamefont {Ong},
  \citenamefont {Fedorov}, \citenamefont {Lin}, \citenamefont {Bansil},
  \citenamefont {Grauer}, \citenamefont {Hor}, \citenamefont {Cava},\ and\
  \citenamefont {Hasan}}]{Hsieh:2009}%
  \BibitemOpen
  \bibfield  {author} {\bibinfo {author} {\bibfnamefont {D.}~\bibnamefont
  {Hsieh}}, \bibinfo {author} {\bibfnamefont {Y.}~\bibnamefont {Xia}}, \bibinfo
  {author} {\bibfnamefont {D.}~\bibnamefont {Qian}}, \bibinfo {author}
  {\bibfnamefont {L.}~\bibnamefont {Wray}}, \bibinfo {author} {\bibfnamefont
  {J.~H.}\ \bibnamefont {Dil}}, \bibinfo {author} {\bibfnamefont
  {F.}~\bibnamefont {Meier}}, \bibinfo {author} {\bibfnamefont
  {J.}~\bibnamefont {Osterwalder}}, \bibinfo {author} {\bibfnamefont
  {L.}~\bibnamefont {Patthey}}, \bibinfo {author} {\bibfnamefont {J.~G.}\
  \bibnamefont {Checkelsky}}, \bibinfo {author} {\bibfnamefont {N.~P.}\
  \bibnamefont {Ong}}, \bibinfo {author} {\bibfnamefont {A.~V.}\ \bibnamefont
  {Fedorov}}, \bibinfo {author} {\bibfnamefont {H.}~\bibnamefont {Lin}},
  \bibinfo {author} {\bibfnamefont {A.}~\bibnamefont {Bansil}}, \bibinfo
  {author} {\bibfnamefont {D.}~\bibnamefont {Grauer}}, \bibinfo {author}
  {\bibfnamefont {Y.~S.}\ \bibnamefont {Hor}}, \bibinfo {author} {\bibfnamefont
  {R.~J.}\ \bibnamefont {Cava}}, \ and\ \bibinfo {author} {\bibfnamefont
  {M.~Z.}\ \bibnamefont {Hasan}},\ }\href {\doibase M3 - 10.1038/nature08234}
  {\bibfield  {journal} {\bibinfo  {journal} {Nature}\ }\textbf {\bibinfo
  {volume} {460}},\ \bibinfo {pages} {1101} (\bibinfo {year}
  {2009})}\BibitemShut {NoStop}%
\bibitem [{\citenamefont {Roushan}\ \emph {et~al.}(2009)\citenamefont
  {Roushan}, \citenamefont {Seo}, \citenamefont {Parker}, \citenamefont {Hor},
  \citenamefont {Hsieh}, \citenamefont {Qian}, \citenamefont {Richardella},
  \citenamefont {Hasan}, \citenamefont {Cava},\ and\ \citenamefont
  {Yazdani}}]{Roushan:2009}%
  \BibitemOpen
  \bibfield  {author} {\bibinfo {author} {\bibfnamefont {P.}~\bibnamefont
  {Roushan}}, \bibinfo {author} {\bibfnamefont {J.}~\bibnamefont {Seo}},
  \bibinfo {author} {\bibfnamefont {C.~V.}\ \bibnamefont {Parker}}, \bibinfo
  {author} {\bibfnamefont {Y.~S.}\ \bibnamefont {Hor}}, \bibinfo {author}
  {\bibfnamefont {D.}~\bibnamefont {Hsieh}}, \bibinfo {author} {\bibfnamefont
  {D.}~\bibnamefont {Qian}}, \bibinfo {author} {\bibfnamefont {A.}~\bibnamefont
  {Richardella}}, \bibinfo {author} {\bibfnamefont {M.~Z.}\ \bibnamefont
  {Hasan}}, \bibinfo {author} {\bibfnamefont {R.~J.}\ \bibnamefont {Cava}}, \
  and\ \bibinfo {author} {\bibfnamefont {A.}~\bibnamefont {Yazdani}},\ }\href
  {\doibase 10.1126/science.1173034} {\bibfield  {journal} {\bibinfo  {journal}
  {Nature}\ }\textbf {\bibinfo {volume} {460}},\ \bibinfo {pages} {1106}
  (\bibinfo {year} {2009})}\BibitemShut {NoStop}%
\bibitem [{\citenamefont {Alpichshev}\ \emph {et~al.}(2011)\citenamefont
  {Alpichshev}, \citenamefont {Analytis}, \citenamefont {Chu}, \citenamefont
  {Fisher},\ and\ \citenamefont {Kapitulnik}}]{Alpichshev:2011}%
  \BibitemOpen
  \bibfield  {author} {\bibinfo {author} {\bibfnamefont {Z.}~\bibnamefont
  {Alpichshev}}, \bibinfo {author} {\bibfnamefont {J.~G.}\ \bibnamefont
  {Analytis}}, \bibinfo {author} {\bibfnamefont {J.-H.}\ \bibnamefont {Chu}},
  \bibinfo {author} {\bibfnamefont {I.~R.}\ \bibnamefont {Fisher}}, \ and\
  \bibinfo {author} {\bibfnamefont {A.}~\bibnamefont {Kapitulnik}},\ }\href
  {\doibase 10.1103/PhysRevB.84.041104} {\bibfield  {journal} {\bibinfo
  {journal} {Phys. Rev. B}\ }\textbf {\bibinfo {volume} {84}},\ \bibinfo
  {pages} {041104} (\bibinfo {year} {2011})}\BibitemShut {NoStop}%
\bibitem [{\citenamefont {Beidenkopf}\ \emph {et~al.}(2011)\citenamefont
  {Beidenkopf}, \citenamefont {Roushan}, \citenamefont {Seo}, \citenamefont
  {Gorman}, \citenamefont {Drozdov}, \citenamefont {Hor}, \citenamefont
  {Cava},\ and\ \citenamefont {Yazdani}}]{Yazdani11}%
  \BibitemOpen
  \bibfield  {author} {\bibinfo {author} {\bibfnamefont {H.}~\bibnamefont
  {Beidenkopf}}, \bibinfo {author} {\bibfnamefont {P.}~\bibnamefont {Roushan}},
  \bibinfo {author} {\bibfnamefont {J.}~\bibnamefont {Seo}}, \bibinfo {author}
  {\bibfnamefont {L.}~\bibnamefont {Gorman}}, \bibinfo {author} {\bibfnamefont
  {I.}~\bibnamefont {Drozdov}}, \bibinfo {author} {\bibfnamefont {Y.-S.}\
  \bibnamefont {Hor}}, \bibinfo {author} {\bibfnamefont {R.~J.}\ \bibnamefont
  {Cava}}, \ and\ \bibinfo {author} {\bibfnamefont {A.}~\bibnamefont
  {Yazdani}},\ }\href@noop {} {\bibfield  {journal} {\bibinfo  {journal}
  {Nature Physics}\ }\textbf {\bibinfo {volume} {7}},\ \bibinfo {pages} {939}
  (\bibinfo {year} {2011})}\BibitemShut {NoStop}%
\bibitem [{\citenamefont {Analytis}()}]{Analytis12}%
  \BibitemOpen
  \bibfield  {author} {\bibinfo {author} {\bibfnamefont {J.~G.}\ \bibnamefont
  {Analytis}},\ }\href@noop {} {\bibinfo  {journal} {(private communication,
  2012)}\ }\BibitemShut {NoStop}%
\bibitem [{\citenamefont {Butch}\ \emph {et~al.}(2010)\citenamefont {Butch},
  \citenamefont {Kirshenbaum}, \citenamefont {Syers}, \citenamefont {Sushkov},
  \citenamefont {Jenkins}, \citenamefont {Drew},\ and\ \citenamefont
  {Paglione}}]{Butch10}%
  \BibitemOpen
\bibfield  {journal} {  }\bibfield  {author} {\bibinfo {author} {\bibfnamefont
  {N.~P.}\ \bibnamefont {Butch}}, \bibinfo {author} {\bibfnamefont
  {K.}~\bibnamefont {Kirshenbaum}}, \bibinfo {author} {\bibfnamefont
  {P.}~\bibnamefont {Syers}}, \bibinfo {author} {\bibfnamefont {A.~B.}\
  \bibnamefont {Sushkov}}, \bibinfo {author} {\bibfnamefont {G.~S.}\
  \bibnamefont {Jenkins}}, \bibinfo {author} {\bibfnamefont {H.~D.}\
  \bibnamefont {Drew}}, \ and\ \bibinfo {author} {\bibfnamefont
  {J.}~\bibnamefont {Paglione}},\ }\href {\doibase 10.1103/PhysRevB.81.241301}
  {\bibfield  {journal} {\bibinfo  {journal} {Phys. Rev. B}\ }\textbf {\bibinfo
  {volume} {81}},\ \bibinfo {pages} {241301} (\bibinfo {year}
  {2010})}\BibitemShut {NoStop}%
\bibitem [{\citenamefont {Taskin}\ and\ \citenamefont {Ando}(2009)}]{Ando09}%
  \BibitemOpen
  \bibfield  {author} {\bibinfo {author} {\bibfnamefont {A.~A.}\ \bibnamefont
  {Taskin}}\ and\ \bibinfo {author} {\bibfnamefont {Y.}~\bibnamefont {Ando}},\
  }\href {\doibase 10.1103/PhysRevB.80.085303} {\bibfield  {journal} {\bibinfo
  {journal} {Phys. Rev. B}\ }\textbf {\bibinfo {volume} {80}},\ \bibinfo
  {pages} {085303} (\bibinfo {year} {2009})}\BibitemShut {NoStop}%
\bibitem [{\citenamefont {Qu}\ \emph {et~al.}(2010)\citenamefont {Qu},
  \citenamefont {Hor}, \citenamefont {Xiong}, \citenamefont {Cava},\ and\
  \citenamefont {Ong}}]{Ong10QO}%
  \BibitemOpen
  \bibfield  {author} {\bibinfo {author} {\bibfnamefont {D.-X.}\ \bibnamefont
  {Qu}}, \bibinfo {author} {\bibfnamefont {Y.~S.}\ \bibnamefont {Hor}},
  \bibinfo {author} {\bibfnamefont {J.}~\bibnamefont {Xiong}}, \bibinfo
  {author} {\bibfnamefont {R.~J.}\ \bibnamefont {Cava}}, \ and\ \bibinfo
  {author} {\bibfnamefont {N.~P.}\ \bibnamefont {Ong}},\ }\href@noop {}
  {\bibfield  {journal} {\bibinfo  {journal} {Science}\ }\textbf {\bibinfo
  {volume} {329}},\ \bibinfo {pages} {821} (\bibinfo {year}
  {2010})}\BibitemShut {NoStop}%
\bibitem [{\citenamefont {Checkelsky}\ \emph {et~al.}(2011)\citenamefont
  {Checkelsky}, \citenamefont {Hor}, \citenamefont {Cava},\ and\ \citenamefont
  {Ong}}]{Ong10}%
  \BibitemOpen
  \bibfield  {author} {\bibinfo {author} {\bibfnamefont {J.~G.}\ \bibnamefont
  {Checkelsky}}, \bibinfo {author} {\bibfnamefont {Y.~S.}\ \bibnamefont {Hor}},
  \bibinfo {author} {\bibfnamefont {R.~J.}\ \bibnamefont {Cava}}, \ and\
  \bibinfo {author} {\bibfnamefont {N.~P.}\ \bibnamefont {Ong}},\ }\href@noop
  {} {\bibfield  {journal} {\bibinfo  {journal} {Phys. Rev. Lett.}\ }\textbf
  {\bibinfo {volume} {106}},\ \bibinfo {pages} {196801} (\bibinfo {year}
  {2011})}\BibitemShut {NoStop}%
\bibitem [{\citenamefont {Analytis}\ \emph
  {et~al.}(2010{\natexlab{b}})\citenamefont {Analytis}, \citenamefont {Chu},
  \citenamefont {Chen}, \citenamefont {Corredor}, \citenamefont {McDonald},
  \citenamefont {Shen},\ and\ \citenamefont {Fisher}}]{Analytis10}%
  \BibitemOpen
  \bibfield  {author} {\bibinfo {author} {\bibfnamefont {J.~G.}\ \bibnamefont
  {Analytis}}, \bibinfo {author} {\bibfnamefont {J.-H.}\ \bibnamefont {Chu}},
  \bibinfo {author} {\bibfnamefont {Y.}~\bibnamefont {Chen}}, \bibinfo {author}
  {\bibfnamefont {F.}~\bibnamefont {Corredor}}, \bibinfo {author}
  {\bibfnamefont {R.~D.}\ \bibnamefont {McDonald}}, \bibinfo {author}
  {\bibfnamefont {Z.~X.}\ \bibnamefont {Shen}}, \ and\ \bibinfo {author}
  {\bibfnamefont {I.~R.}\ \bibnamefont {Fisher}},\ }\href {\doibase
  10.1103/PhysRevB.81.205407} {\bibfield  {journal} {\bibinfo  {journal} {Phys.
  Rev. B}\ }\textbf {\bibinfo {volume} {81}},\ \bibinfo {pages} {205407}
  (\bibinfo {year} {2010}{\natexlab{b}})}\BibitemShut {NoStop}%
\bibitem [{\citenamefont {Steinberg}\ \emph {et~al.}(2010)\citenamefont
  {Steinberg}, \citenamefont {Gardner}, \citenamefont {Lee},\ and\
  \citenamefont {Jarillo-Herrero}}]{Steinberg:2010}%
  \BibitemOpen
  \bibfield  {author} {\bibinfo {author} {\bibfnamefont {H.}~\bibnamefont
  {Steinberg}}, \bibinfo {author} {\bibfnamefont {D.~R.}\ \bibnamefont
  {Gardner}}, \bibinfo {author} {\bibfnamefont {Y.~S.}\ \bibnamefont {Lee}}, \
  and\ \bibinfo {author} {\bibfnamefont {P.}~\bibnamefont {Jarillo-Herrero}},\
  }\href {\doibase 10.1021/nl1032183} {\bibfield  {journal} {\bibinfo
  {journal} {Nano Letters}\ }\textbf {\bibinfo {volume} {10}},\ \bibinfo
  {pages} {5032} (\bibinfo {year} {2010})}\BibitemShut {NoStop}%
\bibitem [{\citenamefont {Hsieh}\ \emph {et~al.}(2011)\citenamefont {Hsieh},
  \citenamefont {Mahmood}, \citenamefont {McIver}, \citenamefont {Gardner},
  \citenamefont {Lee},\ and\ \citenamefont {Gedik}}]{Gedik11}%
  \BibitemOpen
  \bibfield  {author} {\bibinfo {author} {\bibfnamefont {D.}~\bibnamefont
  {Hsieh}}, \bibinfo {author} {\bibfnamefont {F.}~\bibnamefont {Mahmood}},
  \bibinfo {author} {\bibfnamefont {J.~W.}\ \bibnamefont {McIver}}, \bibinfo
  {author} {\bibfnamefont {D.~R.}\ \bibnamefont {Gardner}}, \bibinfo {author}
  {\bibfnamefont {Y.~S.}\ \bibnamefont {Lee}}, \ and\ \bibinfo {author}
  {\bibfnamefont {N.}~\bibnamefont {Gedik}},\ }\href {\doibase
  10.1103/PhysRevLett.107.077401} {\bibfield  {journal} {\bibinfo  {journal}
  {Phys. Rev. Lett.}\ }\textbf {\bibinfo {volume} {107}},\ \bibinfo {pages}
  {077401} (\bibinfo {year} {2011})}\BibitemShut {NoStop}%
\bibitem [{\citenamefont {Kong}\ \emph {et~al.}(2011)\citenamefont {Kong},
  \citenamefont {Cha}, \citenamefont {Lai}, \citenamefont {Peng}, \citenamefont
  {Analytis}, \citenamefont {Meister}, \citenamefont {Chen}, \citenamefont
  {Zhang}, \citenamefont {Fisher}, \citenamefont {Shen},\ and\ \citenamefont
  {Cui}}]{Cui11}%
  \BibitemOpen
  \bibfield  {author} {\bibinfo {author} {\bibfnamefont {D.}~\bibnamefont
  {Kong}}, \bibinfo {author} {\bibfnamefont {J.~J.}\ \bibnamefont {Cha}},
  \bibinfo {author} {\bibfnamefont {K.}~\bibnamefont {Lai}}, \bibinfo {author}
  {\bibfnamefont {H.}~\bibnamefont {Peng}}, \bibinfo {author} {\bibfnamefont
  {J.~G.}\ \bibnamefont {Analytis}}, \bibinfo {author} {\bibfnamefont
  {S.}~\bibnamefont {Meister}}, \bibinfo {author} {\bibfnamefont
  {Y.}~\bibnamefont {Chen}}, \bibinfo {author} {\bibfnamefont {H.-J.}\
  \bibnamefont {Zhang}}, \bibinfo {author} {\bibfnamefont {I.~R.}\ \bibnamefont
  {Fisher}}, \bibinfo {author} {\bibfnamefont {Z.-X.}\ \bibnamefont {Shen}}, \
  and\ \bibinfo {author} {\bibfnamefont {Y.}~\bibnamefont {Cui}},\ }\href@noop
  {} {\bibfield  {journal} {\bibinfo  {journal} {ACS Nano}\ }\textbf {\bibinfo
  {volume} {5}},\ \bibinfo {pages} {4698} (\bibinfo {year} {2011})}\BibitemShut
  {NoStop}%
\bibitem [{\citenamefont {{Vald{\'e}s Aguilar}}\ \emph {et~al.}()\citenamefont
  {{Vald{\'e}s Aguilar}}, \citenamefont {{Wu}}, \citenamefont {{Stier}},
  \citenamefont {{Bilbro}}, \citenamefont {{Brahlek}}, \citenamefont
  {{Bansal}}, \citenamefont {{Oh}},\ and\ \citenamefont
  {{Armitage}}}]{Armitage12}%
  \BibitemOpen
  \bibfield  {author} {\bibinfo {author} {\bibfnamefont {R.}~\bibnamefont
  {{Vald{\'e}s Aguilar}}}, \bibinfo {author} {\bibfnamefont {L.}~\bibnamefont
  {{Wu}}}, \bibinfo {author} {\bibfnamefont {A.~V.}\ \bibnamefont {{Stier}}},
  \bibinfo {author} {\bibfnamefont {L.~S.}\ \bibnamefont {{Bilbro}}}, \bibinfo
  {author} {\bibfnamefont {M.}~\bibnamefont {{Brahlek}}}, \bibinfo {author}
  {\bibfnamefont {N.}~\bibnamefont {{Bansal}}}, \bibinfo {author}
  {\bibfnamefont {S.}~\bibnamefont {{Oh}}}, \ and\ \bibinfo {author}
  {\bibfnamefont {N.~P.}\ \bibnamefont {{Armitage}}},\ }\href@noop {} {\
  }\Eprint {http://arxiv.org/abs/arXiv:1202.1249} {arXiv:1202.1249}
  \BibitemShut {NoStop}%
\bibitem [{\citenamefont {Wray}\ \emph {et~al.}(2011)\citenamefont {Wray},
  \citenamefont {Xu}, \citenamefont {Xia}, \citenamefont {Hsieh}, \citenamefont
  {Fedorov}, \citenamefont {Hor}, \citenamefont {Cava}, \citenamefont {Bansil},
  \citenamefont {Lin},\ and\ \citenamefont {Hasan}}]{Hasan11}%
  \BibitemOpen
  \bibfield  {author} {\bibinfo {author} {\bibfnamefont {L.~A.}\ \bibnamefont
  {Wray}}, \bibinfo {author} {\bibfnamefont {S.-Y.}\ \bibnamefont {Xu}},
  \bibinfo {author} {\bibfnamefont {Y.}~\bibnamefont {Xia}}, \bibinfo {author}
  {\bibfnamefont {D.}~\bibnamefont {Hsieh}}, \bibinfo {author} {\bibfnamefont
  {A.~V.}\ \bibnamefont {Fedorov}}, \bibinfo {author} {\bibfnamefont {Y.~S.}\
  \bibnamefont {Hor}}, \bibinfo {author} {\bibfnamefont {R.~J.}\ \bibnamefont
  {Cava}}, \bibinfo {author} {\bibfnamefont {A.}~\bibnamefont {Bansil}},
  \bibinfo {author} {\bibfnamefont {H.}~\bibnamefont {Lin}}, \ and\ \bibinfo
  {author} {\bibfnamefont {M.~Z.}\ \bibnamefont {Hasan}},\ }\href@noop {}
  {\bibfield  {journal} {\bibinfo  {journal} {Nature Physics}\ }\textbf
  {\bibinfo {volume} {7}},\ \bibinfo {pages} {32} (\bibinfo {year}
  {2011})}\BibitemShut {NoStop}%
\bibitem [{\citenamefont {Hor}\ \emph {et~al.}(2010)\citenamefont {Hor},
  \citenamefont {Williams}, \citenamefont {Checkelsky}, \citenamefont
  {Roushan}, \citenamefont {Seo}, \citenamefont {Xu}, \citenamefont
  {Zandbergen}, \citenamefont {Yazdani}, \citenamefont {Ong},\ and\
  \citenamefont {Cava}}]{Cava10}%
  \BibitemOpen
  \bibfield  {author} {\bibinfo {author} {\bibfnamefont {Y.~S.}\ \bibnamefont
  {Hor}}, \bibinfo {author} {\bibfnamefont {A.~J.}\ \bibnamefont {Williams}},
  \bibinfo {author} {\bibfnamefont {J.~G.}\ \bibnamefont {Checkelsky}},
  \bibinfo {author} {\bibfnamefont {P.}~\bibnamefont {Roushan}}, \bibinfo
  {author} {\bibfnamefont {J.}~\bibnamefont {Seo}}, \bibinfo {author}
  {\bibfnamefont {Q.}~\bibnamefont {Xu}}, \bibinfo {author} {\bibfnamefont
  {H.~W.}\ \bibnamefont {Zandbergen}}, \bibinfo {author} {\bibfnamefont
  {A.}~\bibnamefont {Yazdani}}, \bibinfo {author} {\bibfnamefont {N.~P.}\
  \bibnamefont {Ong}}, \ and\ \bibinfo {author} {\bibfnamefont {R.~J.}\
  \bibnamefont {Cava}},\ }\href {\doibase 10.1103/PhysRevLett.104.057001}
  {\bibfield  {journal} {\bibinfo  {journal} {Phys. Rev. Lett.}\ }\textbf
  {\bibinfo {volume} {104}},\ \bibinfo {pages} {057001} (\bibinfo {year}
  {2010})}\BibitemShut {NoStop}%
\bibitem [{\citenamefont {Sasaki}\ \emph {et~al.}(2011)\citenamefont {Sasaki},
  \citenamefont {Kriener}, \citenamefont {Segawa}, \citenamefont {Yada},
  \citenamefont {Tanaka}, \citenamefont {Sato},\ and\ \citenamefont
  {Ando}}]{Ando11}%
  \BibitemOpen
  \bibfield  {author} {\bibinfo {author} {\bibfnamefont {S.}~\bibnamefont
  {Sasaki}}, \bibinfo {author} {\bibfnamefont {M.}~\bibnamefont {Kriener}},
  \bibinfo {author} {\bibfnamefont {K.}~\bibnamefont {Segawa}}, \bibinfo
  {author} {\bibfnamefont {K.}~\bibnamefont {Yada}}, \bibinfo {author}
  {\bibfnamefont {Y.}~\bibnamefont {Tanaka}}, \bibinfo {author} {\bibfnamefont
  {M.}~\bibnamefont {Sato}}, \ and\ \bibinfo {author} {\bibfnamefont
  {Y.}~\bibnamefont {Ando}},\ }\href@noop {} {\bibfield  {journal} {\bibinfo
  {journal} {Phys. Rev. Lett.}\ }\textbf {\bibinfo {volume} {107}},\ \bibinfo
  {pages} {217001} (\bibinfo {year} {2011})}\BibitemShut {NoStop}%
\bibitem [{\citenamefont {Sinuco-Le\'on}\ \emph {et~al.}(2011)\citenamefont
  {Sinuco-Le\'on}, \citenamefont {Kaczmarek}, \citenamefont {Kr\"uger},\ and\
  \citenamefont {Fromhold}}]{Fromhold11}%
  \BibitemOpen
  \bibfield  {author} {\bibinfo {author} {\bibfnamefont {G.}~\bibnamefont
  {Sinuco-Le\'on}}, \bibinfo {author} {\bibfnamefont {B.}~\bibnamefont
  {Kaczmarek}}, \bibinfo {author} {\bibfnamefont {P.}~\bibnamefont {Kr\"uger}},
  \ and\ \bibinfo {author} {\bibfnamefont {T.~M.}\ \bibnamefont {Fromhold}},\
  }\href@noop {} {\bibfield  {journal} {\bibinfo  {journal} {Phys. Rev. A}\
  }\textbf {\bibinfo {volume} {83}},\ \bibinfo {pages} {021401} (\bibinfo
  {year} {2011})}\BibitemShut {NoStop}%
\bibitem [{\citenamefont {Wildermuth}\ \emph {et~al.}(2005)\citenamefont
  {Wildermuth}, \citenamefont {Hofferberth}, \citenamefont {Lesanovsky},
  \citenamefont {Haller}, \citenamefont {Andersson}, \citenamefont {Groth},
  \citenamefont {Bar-Joseph}, \citenamefont {{Kr\"{u}ger}},\ and\ \citenamefont
  {Schmiedmayer}}]{Schmiedmayer05_Nature}%
  \BibitemOpen
  \bibfield  {author} {\bibinfo {author} {\bibfnamefont {S.}~\bibnamefont
  {Wildermuth}}, \bibinfo {author} {\bibfnamefont {S.}~\bibnamefont
  {Hofferberth}}, \bibinfo {author} {\bibfnamefont {I.}~\bibnamefont
  {Lesanovsky}}, \bibinfo {author} {\bibfnamefont {E.}~\bibnamefont {Haller}},
  \bibinfo {author} {\bibfnamefont {L.~M.}\ \bibnamefont {Andersson}}, \bibinfo
  {author} {\bibfnamefont {S.}~\bibnamefont {Groth}}, \bibinfo {author}
  {\bibfnamefont {I.}~\bibnamefont {Bar-Joseph}}, \bibinfo {author}
  {\bibfnamefont {P.}~\bibnamefont {{Kr\"{u}ger}}}, \ and\ \bibinfo {author}
  {\bibfnamefont {J.}~\bibnamefont {Schmiedmayer}},\ }\href@noop {} {\bibfield
  {journal} {\bibinfo  {journal} {Nature}\ }\textbf {\bibinfo {volume}
  {\textbf{435}}},\ \bibinfo {pages} {440} (\bibinfo {year}
  {2005})}\BibitemShut {NoStop}%
\bibitem [{\citenamefont {Wildermuth}\ \emph {et~al.}(2006)\citenamefont
  {Wildermuth}, \citenamefont {Hofferberth}, \citenamefont {Lesanovsky},
  \citenamefont {Groth}, \citenamefont {{Kr\"{u}ger}}, \citenamefont
  {Schmiedmayer},\ and\ \citenamefont {Bar-Joseph}}]{Schmiedmayer06_APL}%
  \BibitemOpen
  \bibfield  {author} {\bibinfo {author} {\bibfnamefont {S.}~\bibnamefont
  {Wildermuth}}, \bibinfo {author} {\bibfnamefont {S.}~\bibnamefont
  {Hofferberth}}, \bibinfo {author} {\bibfnamefont {I.}~\bibnamefont
  {Lesanovsky}}, \bibinfo {author} {\bibfnamefont {S.}~\bibnamefont {Groth}},
  \bibinfo {author} {\bibfnamefont {P.}~\bibnamefont {{Kr\"{u}ger}}}, \bibinfo
  {author} {\bibfnamefont {J.}~\bibnamefont {Schmiedmayer}}, \ and\ \bibinfo
  {author} {\bibfnamefont {I.}~\bibnamefont {Bar-Joseph}},\ }\href@noop {}
  {\bibfield  {journal} {\bibinfo  {journal} {Appl. Phys. Lett.}\ }\textbf
  {\bibinfo {volume} {\textbf{88}}},\ \bibinfo {pages} {264103} (\bibinfo
  {year} {2006})}\BibitemShut {NoStop}%
\bibitem [{\citenamefont {Aigner}\ \emph {et~al.}(2008)\citenamefont {Aigner},
  \citenamefont {Pietra}, \citenamefont {Japha}, \citenamefont {Entin-Wohlman},
  \citenamefont {David}, \citenamefont {Salem}, \citenamefont {Folman},\ and\
  \citenamefont {Schmiedmayer}}]{Aigner:2008}%
  \BibitemOpen
  \bibfield  {author} {\bibinfo {author} {\bibfnamefont {S.}~\bibnamefont
  {Aigner}}, \bibinfo {author} {\bibfnamefont {L.~D.}\ \bibnamefont {Pietra}},
  \bibinfo {author} {\bibfnamefont {Y.}~\bibnamefont {Japha}}, \bibinfo
  {author} {\bibfnamefont {O.}~\bibnamefont {Entin-Wohlman}}, \bibinfo {author}
  {\bibfnamefont {T.}~\bibnamefont {David}}, \bibinfo {author} {\bibfnamefont
  {R.}~\bibnamefont {Salem}}, \bibinfo {author} {\bibfnamefont
  {R.}~\bibnamefont {Folman}}, \ and\ \bibinfo {author} {\bibfnamefont
  {J.}~\bibnamefont {Schmiedmayer}},\ }\href {\doibase 10.1126/science.1152458}
  {\bibfield  {journal} {\bibinfo  {journal} {Science}\ }\textbf {\bibinfo
  {volume} {319}},\ \bibinfo {pages} {1226} (\bibinfo {year}
  {2008})}\BibitemShut {NoStop}%
\bibitem [{\citenamefont {Reichel}\ \emph {et~al.}(2001)\citenamefont
  {Reichel}, \citenamefont {H\"{a}nsel}, \citenamefont {Hommelhoff},\ and\
  \citenamefont {H\"{a}nsch}}]{Reichel01a}%
  \BibitemOpen
  \bibfield  {author} {\bibinfo {author} {\bibfnamefont {J.}~\bibnamefont
  {Reichel}}, \bibinfo {author} {\bibfnamefont {W.}~\bibnamefont {H\"{a}nsel}},
  \bibinfo {author} {\bibfnamefont {P.}~\bibnamefont {Hommelhoff}}, \ and\
  \bibinfo {author} {\bibfnamefont {T.~W.}\ \bibnamefont {H\"{a}nsch}},\
  }\href@noop {} {\bibfield  {journal} {\bibinfo  {journal} {Appl. Phys. B}\
  }\textbf {\bibinfo {volume} {\textbf{72}}},\ \bibinfo {pages} {81} (\bibinfo
  {year} {2001})}\BibitemShut {NoStop}%
\bibitem [{\citenamefont {Folman}\ \emph {et~al.}(2002)\citenamefont {Folman},
  \citenamefont {Kr\"{u}ger}, \citenamefont {Schmiedmayer}, \citenamefont
  {Denschlag},\ and\ \citenamefont {Henkel}}]{Schmiedmayer02}%
  \BibitemOpen
  \bibfield  {author} {\bibinfo {author} {\bibfnamefont {R.}~\bibnamefont
  {Folman}}, \bibinfo {author} {\bibfnamefont {P.}~\bibnamefont {Kr\"{u}ger}},
  \bibinfo {author} {\bibfnamefont {J.}~\bibnamefont {Schmiedmayer}}, \bibinfo
  {author} {\bibfnamefont {J.}~\bibnamefont {Denschlag}}, \ and\ \bibinfo
  {author} {\bibfnamefont {C.}~\bibnamefont {Henkel}},\ }\href@noop {}
  {\bibfield  {journal} {\bibinfo  {journal} {Adv. At. Mol. Opt. Phys.}\
  }\textbf {\bibinfo {volume} {\textbf{48}}},\ \bibinfo {pages} {263} (\bibinfo
  {year} {2002})}\BibitemShut {NoStop}%
\bibitem [{\citenamefont {Fortagh}\ and\ \citenamefont
  {Zimmermann}(2007)}]{Zimmermann07}%
  \BibitemOpen
  \bibfield  {author} {\bibinfo {author} {\bibfnamefont {J.}~\bibnamefont
  {Fortagh}}\ and\ \bibinfo {author} {\bibfnamefont {J.~C.}\ \bibnamefont
  {Zimmermann}},\ }\href@noop {} {\bibfield  {journal} {\bibinfo  {journal}
  {Rev. Mod. Phys.}\ }\textbf {\bibinfo {volume} {\textbf{79}}},\ \bibinfo
  {pages} {235} (\bibinfo {year} {2007})}\BibitemShut {NoStop}%
\bibitem [{\citenamefont {Kr{\"u}ger}\ \emph {et~al.}(2007)\citenamefont
  {Kr{\"u}ger}, \citenamefont {Andersson}, \citenamefont {Wildermuth},
  \citenamefont {Hofferberth}, \citenamefont {Haller}, \citenamefont {Aigner},
  \citenamefont {Groth}, \citenamefont {Bar-Joseph},\ and\ \citenamefont
  {Schmiedmayer}}]{Kruger07}%
  \BibitemOpen
  \bibfield  {author} {\bibinfo {author} {\bibfnamefont {P.}~\bibnamefont
  {Kr{\"u}ger}}, \bibinfo {author} {\bibfnamefont {L.~M.}\ \bibnamefont
  {Andersson}}, \bibinfo {author} {\bibfnamefont {S.}~\bibnamefont
  {Wildermuth}}, \bibinfo {author} {\bibfnamefont {S.}~\bibnamefont
  {Hofferberth}}, \bibinfo {author} {\bibfnamefont {E.}~\bibnamefont {Haller}},
  \bibinfo {author} {\bibfnamefont {S.}~\bibnamefont {Aigner}}, \bibinfo
  {author} {\bibfnamefont {S.}~\bibnamefont {Groth}}, \bibinfo {author}
  {\bibfnamefont {I.}~\bibnamefont {Bar-Joseph}}, \ and\ \bibinfo {author}
  {\bibfnamefont {J.}~\bibnamefont {Schmiedmayer}},\ }\href@noop {} {\bibfield
  {journal} {\bibinfo  {journal} {Phys. Rev. A}\ }\textbf {\bibinfo {volume}
  {76}},\ \bibinfo {pages} {063621} (\bibinfo {year} {2007})}\BibitemShut
  {NoStop}%
\bibitem [{\citenamefont {Weinstein}\ and\ \citenamefont
  {Libbrecht}(1995)}]{Libbrecht95a}%
  \BibitemOpen
  \bibfield  {author} {\bibinfo {author} {\bibfnamefont {J.~D.}\ \bibnamefont
  {Weinstein}}\ and\ \bibinfo {author} {\bibfnamefont {K.~G.}\ \bibnamefont
  {Libbrecht}},\ }\href@noop {} {\bibfield  {journal} {\bibinfo  {journal}
  {Phys. Rev. A}\ }\textbf {\bibinfo {volume} {\textbf{52}}},\ \bibinfo {pages}
  {4004} (\bibinfo {year} {1995})}\BibitemShut {NoStop}%
\bibitem [{\citenamefont {Lin}\ \emph {et~al.}(2004)\citenamefont {Lin},
  \citenamefont {Teper}, \citenamefont {Chin},\ and\ \citenamefont
  {{Vuleti\'{c}}}}]{Vladan04}%
  \BibitemOpen
  \bibfield  {author} {\bibinfo {author} {\bibfnamefont {Y.-J.}\ \bibnamefont
  {Lin}}, \bibinfo {author} {\bibfnamefont {I.}~\bibnamefont {Teper}}, \bibinfo
  {author} {\bibfnamefont {C.}~\bibnamefont {Chin}}, \ and\ \bibinfo {author}
  {\bibfnamefont {V.}~\bibnamefont {{Vuleti\'{c}}}},\ }\href@noop {} {\bibfield
   {journal} {\bibinfo  {journal} {Phys. Rev. Lett.}\ }\textbf {\bibinfo
  {volume} {92}},\ \bibinfo {pages} {050404} (\bibinfo {year}
  {2004})}\BibitemShut {NoStop}%
\bibitem [{\citenamefont {Kr\"{u}ger}()}]{Kruger04}%
  \BibitemOpen
  \bibfield  {author} {\bibinfo {author} {\bibfnamefont {P.}~\bibnamefont
  {Kr\"{u}ger}},\ }\href@noop {} {\bibinfo  {journal} {Ph.D. thesis, University
  of Heidelberg, (2004)}\ }\BibitemShut {NoStop}%
\bibitem [{\citenamefont {{Kr\"{u}ger}}\ \emph {et~al.}(2005)\citenamefont
  {{Kr\"{u}ger}}, \citenamefont {Wildermuth}, \citenamefont {Hofferberth},
  \citenamefont {Andersson}, \citenamefont {Groth}, \citenamefont
  {Bar-Joseph},\ and\ \citenamefont {Schmiedmayer}}]{Kruger05}%
  \BibitemOpen
\bibfield  {journal} {  }\bibfield  {author} {\bibinfo {author} {\bibfnamefont
  {P.}~\bibnamefont {{Kr\"{u}ger}}}, \bibinfo {author} {\bibfnamefont
  {S.}~\bibnamefont {Wildermuth}}, \bibinfo {author} {\bibfnamefont
  {S.}~\bibnamefont {Hofferberth}}, \bibinfo {author} {\bibfnamefont {L.~M.}\
  \bibnamefont {Andersson}}, \bibinfo {author} {\bibfnamefont {S.}~\bibnamefont
  {Groth}}, \bibinfo {author} {\bibfnamefont {I.}~\bibnamefont {Bar-Joseph}}, \
  and\ \bibinfo {author} {\bibfnamefont {J.}~\bibnamefont {Schmiedmayer}},\
  }\href@noop {} {\bibfield  {journal} {\bibinfo  {journal} {Journal of
  Physics: Conference Series}\ }\textbf {\bibinfo {volume} {19}} (\bibinfo
  {year} {2005})}\BibitemShut {NoStop}%
\bibitem [{\citenamefont {Smith}\ \emph {et~al.}(2011)\citenamefont {Smith},
  \citenamefont {Aigner}, \citenamefont {Hofferberth}, \citenamefont {Gring},
  \citenamefont {Andersson}, \citenamefont {Wildermuth}, \citenamefont
  {{Kr\"{u}ger}}, \citenamefont {Schneider}, \citenamefont {Schumm},\ and\
  \citenamefont {Schmiedmayer}}]{Schmiedmayer11}%
  \BibitemOpen
  \bibfield  {author} {\bibinfo {author} {\bibfnamefont {D.~A.}\ \bibnamefont
  {Smith}}, \bibinfo {author} {\bibfnamefont {S.}~\bibnamefont {Aigner}},
  \bibinfo {author} {\bibfnamefont {S.}~\bibnamefont {Hofferberth}}, \bibinfo
  {author} {\bibfnamefont {M.}~\bibnamefont {Gring}}, \bibinfo {author}
  {\bibfnamefont {M.}~\bibnamefont {Andersson}}, \bibinfo {author}
  {\bibfnamefont {S.}~\bibnamefont {Wildermuth}}, \bibinfo {author}
  {\bibfnamefont {P.}~\bibnamefont {{Kr\"{u}ger}}}, \bibinfo {author}
  {\bibfnamefont {S.}~\bibnamefont {Schneider}}, \bibinfo {author}
  {\bibfnamefont {T.}~\bibnamefont {Schumm}}, \ and\ \bibinfo {author}
  {\bibfnamefont {J.}~\bibnamefont {Schmiedmayer}},\ }\href@noop {} {\bibfield
  {journal} {\bibinfo  {journal} {Opt. Express}\ }\textbf {\bibinfo {volume}
  {19}},\ \bibinfo {pages} {8471} (\bibinfo {year} {2011})}\BibitemShut
  {NoStop}%
\bibitem [{\citenamefont {Henkel}\ \emph {et~al.}(1999)\citenamefont {Henkel},
  \citenamefont {{P\"{o}tting}},\ and\ \citenamefont {Wilkens}}]{Henkel99}%
  \BibitemOpen
  \bibfield  {author} {\bibinfo {author} {\bibfnamefont {C.}~\bibnamefont
  {Henkel}}, \bibinfo {author} {\bibfnamefont {S.}~\bibnamefont
  {{P\"{o}tting}}}, \ and\ \bibinfo {author} {\bibfnamefont {M.}~\bibnamefont
  {Wilkens}},\ }\href@noop {} {\bibfield  {journal} {\bibinfo  {journal} {Appl.
  Phys. B}\ }\textbf {\bibinfo {volume} {69}},\ \bibinfo {pages} {379}
  (\bibinfo {year} {1999})}\BibitemShut {NoStop}%
\bibitem [{\citenamefont {Henkel}\ and\ \citenamefont
  {{P\"{o}tting}}(2001)}]{Henkel01}%
  \BibitemOpen
  \bibfield  {author} {\bibinfo {author} {\bibfnamefont {C.}~\bibnamefont
  {Henkel}}\ and\ \bibinfo {author} {\bibfnamefont {S.}~\bibnamefont
  {{P\"{o}tting}}},\ }\href@noop {} {\bibfield  {journal} {\bibinfo  {journal}
  {Appl. Phys. B}\ }\textbf {\bibinfo {volume} {72}},\ \bibinfo {pages} {73}
  (\bibinfo {year} {2001})}\BibitemShut {NoStop}%
\bibitem [{\citenamefont {Zimmermann}\ \emph {et~al.}(2011)\citenamefont
  {Zimmermann}, \citenamefont {{M\"{u}ller}}, \citenamefont {Meineke},
  \citenamefont {Esslinger},\ and\ \citenamefont {Moritz}}]{Esslinger11}%
  \BibitemOpen
  \bibfield  {author} {\bibinfo {author} {\bibfnamefont {B.}~\bibnamefont
  {Zimmermann}}, \bibinfo {author} {\bibfnamefont {T.}~\bibnamefont
  {{M\"{u}ller}}}, \bibinfo {author} {\bibfnamefont {J.}~\bibnamefont
  {Meineke}}, \bibinfo {author} {\bibfnamefont {T.}~\bibnamefont {Esslinger}},
  \ and\ \bibinfo {author} {\bibfnamefont {H.}~\bibnamefont {Moritz}},\
  }\href@noop {} {\bibfield  {journal} {\bibinfo  {journal} {New Journal of
  Physics}\ }\textbf {\bibinfo {volume} {13}},\ \bibinfo {pages} {043007}
  (\bibinfo {year} {2011})}\BibitemShut {NoStop}%
\bibitem [{\citenamefont {McGuirk}\ \emph {et~al.}(2004)\citenamefont
  {McGuirk}, \citenamefont {Harber}, \citenamefont {Obrecht},\ and\
  \citenamefont {Cornell}}]{Cornell04}%
  \BibitemOpen
  \bibfield  {author} {\bibinfo {author} {\bibfnamefont {J.~M.}\ \bibnamefont
  {McGuirk}}, \bibinfo {author} {\bibfnamefont {D.~M.}\ \bibnamefont {Harber}},
  \bibinfo {author} {\bibfnamefont {J.~M.}\ \bibnamefont {Obrecht}}, \ and\
  \bibinfo {author} {\bibfnamefont {E.~A.}\ \bibnamefont {Cornell}},\
  }\href@noop {} {\bibfield  {journal} {\bibinfo  {journal} {Phys. Rev. A}\
  }\textbf {\bibinfo {volume} {69}},\ \bibinfo {pages} {062905} (\bibinfo
  {year} {2004})}\BibitemShut {NoStop}%
\bibitem [{\citenamefont {Obrecht}\ \emph
  {et~al.}(2007{\natexlab{a}})\citenamefont {Obrecht}, \citenamefont {Wild},
  \citenamefont {Antezza}, \citenamefont {Pitaevskii}, \citenamefont
  {Stringari},\ and\ \citenamefont {Cornell}}]{Cornell07}%
  \BibitemOpen
  \bibfield  {author} {\bibinfo {author} {\bibfnamefont {J.~M.}\ \bibnamefont
  {Obrecht}}, \bibinfo {author} {\bibfnamefont {R.~J.}\ \bibnamefont {Wild}},
  \bibinfo {author} {\bibfnamefont {M.}~\bibnamefont {Antezza}}, \bibinfo
  {author} {\bibfnamefont {L.~P.}\ \bibnamefont {Pitaevskii}}, \bibinfo
  {author} {\bibfnamefont {S.}~\bibnamefont {Stringari}}, \ and\ \bibinfo
  {author} {\bibfnamefont {E.~A.}\ \bibnamefont {Cornell}},\ }\href@noop {}
  {\bibfield  {journal} {\bibinfo  {journal} {Phys. Rev. Lett.}\ }\textbf
  {\bibinfo {volume} {98}},\ \bibinfo {pages} {063201} (\bibinfo {year}
  {2007}{\natexlab{a}})}\BibitemShut {NoStop}%
\bibitem [{\citenamefont {Obrecht}\ \emph
  {et~al.}(2007{\natexlab{b}})\citenamefont {Obrecht}, \citenamefont {Wild},\
  and\ \citenamefont {Cornell}}]{Cornell07b}%
  \BibitemOpen
  \bibfield  {author} {\bibinfo {author} {\bibfnamefont {J.~M.}\ \bibnamefont
  {Obrecht}}, \bibinfo {author} {\bibfnamefont {R.~J.}\ \bibnamefont {Wild}}, \
  and\ \bibinfo {author} {\bibfnamefont {E.~A.}\ \bibnamefont {Cornell}},\
  }\href@noop {} {\bibfield  {journal} {\bibinfo  {journal} {Phys. Rev. A}\
  }\textbf {\bibinfo {volume} {75}},\ \bibinfo {pages} {062903} (\bibinfo
  {year} {2007}{\natexlab{b}})}\BibitemShut {NoStop}%
\bibitem [{\citenamefont {Datta}(2000)}]{Datta00}%
  \BibitemOpen
  \bibfield  {author} {\bibinfo {author} {\bibfnamefont {S.}~\bibnamefont
  {Datta}},\ }\href@noop {} {\bibfield  {journal} {\bibinfo  {journal}
  {Superlattices and Microstructures}\ }\textbf {\bibinfo {volume} {28}},\
  \bibinfo {pages} {253} (\bibinfo {year} {2000})}\BibitemShut {NoStop}%
\bibitem [{\citenamefont {Zhang}\ \emph {et~al.}(2009)\citenamefont {Zhang},
  \citenamefont {Liu}, \citenamefont {Qi}, \citenamefont {Dai}, \citenamefont
  {Fang},\ and\ \citenamefont {Zhang}}]{Zhang:2009}%
  \BibitemOpen
  \bibfield  {author} {\bibinfo {author} {\bibfnamefont {H.}~\bibnamefont
  {Zhang}}, \bibinfo {author} {\bibfnamefont {C.-X.}\ \bibnamefont {Liu}},
  \bibinfo {author} {\bibfnamefont {X.-L.}\ \bibnamefont {Qi}}, \bibinfo
  {author} {\bibfnamefont {X.}~\bibnamefont {Dai}}, \bibinfo {author}
  {\bibfnamefont {Z.}~\bibnamefont {Fang}}, \ and\ \bibinfo {author}
  {\bibfnamefont {S.-C.}\ \bibnamefont {Zhang}},\ }\href {\doibase
  10.1038/nphys1270} {\bibfield  {journal} {\bibinfo  {journal} {Nat Phys}\
  }\textbf {\bibinfo {volume} {5}},\ \bibinfo {pages} {438} (\bibinfo {year}
  {2009})}\BibitemShut {NoStop}%
\bibitem [{\citenamefont {Liu}\ \emph {et~al.}(2010)\citenamefont {Liu},
  \citenamefont {Qi}, \citenamefont {Zhang}, \citenamefont {Dai}, \citenamefont
  {Fang},\ and\ \citenamefont {Zhang}}]{Liu:2010}%
  \BibitemOpen
  \bibfield  {author} {\bibinfo {author} {\bibfnamefont {C.-X.}\ \bibnamefont
  {Liu}}, \bibinfo {author} {\bibfnamefont {X.-L.}\ \bibnamefont {Qi}},
  \bibinfo {author} {\bibfnamefont {H.}~\bibnamefont {Zhang}}, \bibinfo
  {author} {\bibfnamefont {X.}~\bibnamefont {Dai}}, \bibinfo {author}
  {\bibfnamefont {Z.}~\bibnamefont {Fang}}, \ and\ \bibinfo {author}
  {\bibfnamefont {S.-C.}\ \bibnamefont {Zhang}},\ }\href {\doibase
  10.1103/PhysRevB.82.045122} {\bibfield  {journal} {\bibinfo  {journal} {Phys.
  Rev. B}\ }\textbf {\bibinfo {volume} {82}},\ \bibinfo {pages} {045122}
  (\bibinfo {year} {2010})}\BibitemShut {NoStop}%
\bibitem [{\citenamefont {Datta}(2005)}]{SDattaQT}%
  \BibitemOpen
  \bibfield  {author} {\bibinfo {author} {\bibfnamefont {S.}~\bibnamefont
  {Datta}},\ }\href@noop {} {\emph {\bibinfo {title} {Quantum Transport: Atom
  to Transistor}}}\ (\bibinfo  {publisher} {Cambridge University Press},\
  \bibinfo {year} {2005})\BibitemShut {NoStop}%
\bibitem [{Note1()}]{Note1}%
  \BibitemOpen
  \bibinfo {note} {The magnetic susceptibility of Bi$_2$Se$_3$ is less than
  10$^{-6}$ cm$^3$ mol$^{-1}$~\cite {Uemura:1977}.}\BibitemShut {Stop}%
\bibitem [{\citenamefont {Maze}\ \emph {et~al.}(2008)\citenamefont {Maze},
  \citenamefont {Stanwix}, \citenamefont {Hodges}, \citenamefont {Hong},
  \citenamefont {Taylor}, \citenamefont {Cappellaro}, \citenamefont {Jiang},
  \citenamefont {Dutt}, \citenamefont {Togan}, \citenamefont {Zibrov},
  \citenamefont {Yacoby}, \citenamefont {Walsworth},\ and\ \citenamefont
  {Lukin}}]{Lukin08}%
  \BibitemOpen
  \bibfield  {author} {\bibinfo {author} {\bibfnamefont {J.~R.}\ \bibnamefont
  {Maze}}, \bibinfo {author} {\bibfnamefont {P.~L.}\ \bibnamefont {Stanwix}},
  \bibinfo {author} {\bibfnamefont {J.~S.}\ \bibnamefont {Hodges}}, \bibinfo
  {author} {\bibfnamefont {S.}~\bibnamefont {Hong}}, \bibinfo {author}
  {\bibfnamefont {J.~M.}\ \bibnamefont {Taylor}}, \bibinfo {author}
  {\bibfnamefont {P.}~\bibnamefont {Cappellaro}}, \bibinfo {author}
  {\bibfnamefont {L.}~\bibnamefont {Jiang}}, \bibinfo {author} {\bibfnamefont
  {M.~V.~G.}\ \bibnamefont {Dutt}}, \bibinfo {author} {\bibfnamefont
  {E.}~\bibnamefont {Togan}}, \bibinfo {author} {\bibfnamefont {A.~S.}\
  \bibnamefont {Zibrov}}, \bibinfo {author} {\bibfnamefont {A.}~\bibnamefont
  {Yacoby}}, \bibinfo {author} {\bibfnamefont {R.~L.}\ \bibnamefont
  {Walsworth}}, \ and\ \bibinfo {author} {\bibfnamefont {M.~.~D.}\ \bibnamefont
  {Lukin}},\ }\href@noop {} {\bibfield  {journal} {\bibinfo  {journal}
  {Nature}\ }\textbf {\bibinfo {volume} {\textbf{455}}},\ \bibinfo {pages}
  {644} (\bibinfo {year} {2008})}\BibitemShut {NoStop}%
\bibitem [{\citenamefont {Steinert}\ \emph {et~al.}(2010)\citenamefont
  {Steinert}, \citenamefont {Dolde}, \citenamefont {Neumann}, \citenamefont
  {Aird}, \citenamefont {Naydenov}, \citenamefont {Balasubramanian},
  \citenamefont {Jelezko},\ and\ \citenamefont {Wrachtrup}}]{Wrachtrup10}%
  \BibitemOpen
  \bibfield  {author} {\bibinfo {author} {\bibfnamefont {S.}~\bibnamefont
  {Steinert}}, \bibinfo {author} {\bibfnamefont {F.}~\bibnamefont {Dolde}},
  \bibinfo {author} {\bibfnamefont {P.}~\bibnamefont {Neumann}}, \bibinfo
  {author} {\bibfnamefont {A.}~\bibnamefont {Aird}}, \bibinfo {author}
  {\bibfnamefont {B.}~\bibnamefont {Naydenov}}, \bibinfo {author}
  {\bibfnamefont {G.}~\bibnamefont {Balasubramanian}}, \bibinfo {author}
  {\bibfnamefont {F.}~\bibnamefont {Jelezko}}, \ and\ \bibinfo {author}
  {\bibfnamefont {J.}~\bibnamefont {Wrachtrup}},\ }\href@noop {} {\bibfield
  {journal} {\bibinfo  {journal} {Review of Scientific Instruments}\ }\textbf
  {\bibinfo {volume} {81}},\ \bibinfo {pages} {043705} (\bibinfo {year}
  {2010})}\BibitemShut {NoStop}%
\bibitem [{\citenamefont {Uemura}\ and\ \citenamefont
  {Satow}(1977)}]{Uemura:1977}%
  \BibitemOpen
  \bibfield  {author} {\bibinfo {author} {\bibfnamefont {O.}~\bibnamefont
  {Uemura}}\ and\ \bibinfo {author} {\bibfnamefont {T.}~\bibnamefont {Satow}},\
  }\href {\doibase 10.1002/pssb.2220840137} {\bibfield  {journal} {\bibinfo
  {journal} {Physica Status Solidi (b)}\ }\textbf {\bibinfo {volume} {84}},\
  \bibinfo {pages} {353} (\bibinfo {year} {1977})}\BibitemShut {NoStop}%
\end{thebibliography}

%

\end{document}